\documentclass[]{rsproca_new}
\jname{rspa}

\usepackage{tikz,xcolor,hyperref}
\usepackage{amsmath,amsfonts,amssymb}
\usepackage{longtable}
\usepackage{caption}
\usepackage{bm}

 \usepackage{lineno}
 
\definecolor{lime}{HTML}{A6CE39}
\DeclareRobustCommand{\orcidicon}{%
	\begin{tikzpicture}
	\draw[lime, fill=lime] (0,0) 
	circle [radius=0.16] 
	node[white] {{\fontfamily{qag}\selectfont \tiny ID}};
	\draw[white, fill=white] (-0.0625,0.095) 
	circle [radius=0.007];
	\end{tikzpicture}
	\hspace{-2mm}
}

\foreach \x in {A, ..., Z}{%
	\expandafter\xdef\csname orcid\x\endcsname{\noexpand\href{https://orcid.org/\csname orcidauthor\x\endcsname}{\noexpand\orcidicon}}
}

\begin{document}

\title{Characterisation of hydromagnetic waves propagating over a steady, non-axisymmetric background magnetic field}

\author{O. Barrois$^1$, J. Aubert$^1$}

\address{$^1$ Universit\'e de Paris, Institut de Physique du Globe de Paris, CNRS, F-75005 Paris, France.\\}

\subject{geophysics, fluid mechanics, applied mathematics}


\corres{Olivier Barrois\\\email{obarrois@ipgp.fr}}
\esm{}

\begin{abstract}

Motivated by recent observations of rapid (interannual) signals in the geomagnetic data, and by advances in numerical simulations approaching the Earth's outer core conditions, we present a study on the dynamics of hydromagnetic waves evolving over a static base state.
Under the assumption of timescales separation between the rapid waves and the slow convection, we linearise the classical magneto-hydrodynamics equations over a steady non-axisymmetric background magnetic field and a zero velocity field.
The initial perturbation is a super-rotating pulse of the inner core, which sets the amplitude and length-scales of the waves in the system.
The initial pulse triggers axisymmetric, outward propagating torsional Alfv\'en waves, with characteristic thickness scaling with the magnetic Ekman number as $Ek_M^{1/4}$.
Because the background state is non-axisymmetric, the pulse also triggers non-axisymmetric, quasi-geostrophic Alfv\'en waves.
As these latter waves propagate outwards, they turn into quasi-geostrophic, magneto-Coriolis waves (QG-MC) as the Coriolis force supersedes inertia in the force balance.
The period of the initial wave packet is preserved across the shell but the QG-MC wave front disperses and a westward drift is observed after this transformation. 
Upon reaching the core surface, the westward drift of the QG-MC waves presents an estimated phase speed of about $1100\,km/y$.
This analysis confirms the QG-MC nature of the rapid magnetic signals observed in geomagnetic field models near the equator.

\end{abstract}

\begin{fmtext}
\end{fmtext}

\label{firstpage}
\maketitle

\section{Introduction}
\label{sec:state_of_art}

Rapid variations of the Earth's core magnetic field are 
a topic of active current investigation.
Rapid in this context refers to timescales that are much shorter than the overturn time of the core, $\tau_u = d/U_c \sim 125\,y$ -- where $d = 2258.5\,km$ is the thickness of the Earth's outer core and $U_c \sim 20\,km/y$ is a typical convective velocity of the fluid it contains -- and corresponds to variations over a year up to several decades.
One example is the recurrence of geomagnetic jerks, first reported in $1969$-$1970$ \cite{courtillot1978acceleration}, and since regularly detected ({\it e.g.}, \cite{chulliat2010observation,pinheiro2011measurements,brown2013jerks,finlay2020chaos}).
With the improvement of the observational geomagnetic field models ({\it e.g.}, \cite{lesur2010modelling,finlay2016recent}), jerks have been associated with short sequences of pulses of alternating polarities in the magnetic field's second time-derivative \cite{chulliat2010observation}.
The nature of these pulses have frequently been interpreted as waves propagating in the Earth's core, such as magnetic Rossby waves \cite{chulliat2014geomagnetic,chulliat2015fast}.
More recently, the arrival of Alfv\'en waves packets \cite{aubert2019geomagnetic} and Quasi-Geostrophic Magneto-Coriolis waves \cite{aubert2022taxonomy} has been proposed as a source for the jerks \cite{aubert2021interplay}, while it has been shown that such waves with a suitable period could exist in the Earth's outer core \cite{gerick2021fast}.

Waves are fundamentally the result of a restoring force acting within a destabilised system.
They basically arise when the primary balance of a system is disturbed and returns to equilibrium under the influence of a sub-dominant force.
In the outer core of the Earth, it is widely believed that a primary balance holds between the Coriolis force and the pressure, {\it i.e.} the Quasi-Geostrophic balance (hereafter QG) \cite{davidson2013scaling}.
Although this leading-order equilibrium is still debated (see {\it e.g.}, \cite{roberts1965analysis,dormy2016strong}), some of the most advanced Earth-like three-dimensional simulations ({\it e.g.}, \cite{schaeffer2017turbulent,aubert2019approaching}) yield a QG balance at leading order, supplemented by a secondary balance between the Lorentz force, the buoyancy and the remaining of the Coriolis force.
This force balance comprising a Quasi-Geostrophic $0^\mathrm{th}$ order equilibrium and a $1^\mathrm{st}$ order balance between the Magnetic, the Archimedean and the remaining of the Coriolis forces, has been termed a QG-MAC balance ({\it e.g.}, \cite{schwaiger2019force}).
Inertia comes then at second order, and is the next force that can accommodate deviations from the QG-MAC balance because viscosity comes even further below. 
Several types of waves are then possible depending on the restoring balance.
Axisymmetric torsional and quasi-geostrophic Alfv\'en waves \cite{aubert2019geomagnetic}, evolving on annual to decadal timescales, are sustained by a magneto-inertial balance.
Magneto-Coriolis waves \cite{gerick2021fast}, evolving on multi-annual up to millennial timescales, are sustained by a magneto-Coriolis balance.
And 'slow', quasi-geostrophic Rossby waves \cite{zhang2001inertial,canet2014hydromagnetic}, evolving on sub-annual timescales, are finally the result of a Coriolis-inertial balance.
Note that waves restored in part by buoyancy such as MAC waves \cite{buffett2019equatorially} are only possible with stabilising buoyancy, {\it e.g.} within a stratified layer \cite{buffett2019equatorially}.

Advanced three-dimensional simulations (see {\it e.g.}, \cite{aubert2023state}), and improving satellite data \cite{finlay2020chaos,lesur2022rapid} have shown that there is a timescale separation between the fast (interannual) wave dynamics and the slower (secular) convective processes of the dynamo.
Understanding wave dynamics is thus essential for understanding the rapid variations of the magnetic field from the Earth's core.
A clear separation between the different dynamical timescales is then crucial: at times scales longer than the core overturn time, the waves will be disrupted by the convection, and at timescales shorter than the rotation period $\tau_\Omega = 1/\Omega$ -- where $\Omega$ is the rotation period of the Earth --, the effects of rotation are no longer dominant.
Between these daily and centennial timescales, the Alfv\'en time $\tau_{\cal A} = d\,\sqrt{\rho\,\mu}/B_\oplus$ -- about $\sim 2$ years in the core \cite{gillet2010fast}, where $B_\oplus$ is the rms value of the Earth's magnetic field, and $\rho$ and $\mu$ are respectively the density and the magnetic permeability of the Earth's core -- is the fundamental timescale in the vicinity of which most of the wave dynamics is expected to occur.
This timescale is accessible with satellite geomagnetic observations \cite{finlay2020chaos}.
In particular, QG-MC waves \cite{labbe2015magnetostrophic,gerick2021fast} are accessible within the current temporal resolution of the geomagnetic data and have been identified in core flows retrieved from geomagnetic variations \cite{gillet2022satellite}.

However, it is still difficult to clearly separate waves from the convection in three-dimensional simulations, because the spectra around $\tau_u$ and $\tau_{\cal A}$ -- whose values are too close in simulations far from the Earth's core parameters -- overlap.
The rapid dynamics being nonetheless a deviation from the slower convective dynamo and involving the much weaker inertia in the core, it can be seen as a small and linear perturbation about a static base state.
Such small perturbation then allows to linearise the magneto-hydro-dynamics equations around a static background state, inducing a better separation of the different timescales.
A similar strategy was previously followed to study axisymmetric \cite{jault2008axial} and non-axisymmetric \cite{gillet2011rationale} perturbations. Here we follow the same strategy to pursue the study and characterisation of rapid, non-axisymmetric waves in the light of the recent understanding brought by \cite{gerick2021fast,gillet2022satellite}.

The paper is organised as follows.
The methodology is described in \S~\ref{sec:Method}, and the taxonomy of waves that are expected in our setup is detailed in \S~\ref{sec:Waves-types}.
We then present and analyse our results in \S~\ref{sec:Results}, before discussing and proposing some perspectives of this work in \S~\ref{sec:Conclusion}.

\section{Methodology}
\label{sec:Method}

\subsection{Linearised Magneto-hydrodynamic equations}
\label{sec:MHD-lin_equations}

We use a three-dimensional dynamical model to study an incompressible fluid in a thick spherical shell.
The spherical coordinates system $(r, \theta, \phi)$ with unit vectors $({\bm e}_r, {\bm e}_\theta, {\bm e}_\phi)$ is used.
Our model solves for the velocity field ${\bf U}$ and magnetic field ${\bf B}$ of an electrically conducting rapidly rotating spherical shell -- assumed to be at a constant angular velocity ${\bm \Omega} = \Omega {\bm e}_z$.
The shell has a thickness $d = r_o - r_i$ with an aspect ratio $r_i/r_o = 0.35$, suitable for a planetary core such as the Earth's outer core -- where $r_i$ and $r_o$ thus are respectively the radii of the inner core boundary (hereafter ICB) and of the core mantle boundary (hereafter CMB).
A solid inner core is present from $r = 0$ to $r = r_i$.
Electromagnetic conditions are conducting at  $r = r_i$ and insulating at $r = r_o$.
Additionally, the mechanical boundaries are stress-free at both $r_i$ and $r_o$ and the inner core can rotate freely (the kinetic momentum will be conserved for the whole system).

We place ourselves in the setup of \cite{jault2008axial} and linearise our system around a chosen background state magnetic field ${\bf B}_0$ and an arbitrary background velocity field ${\bf U}_0$, {\it i.e.} ${\bf U} \equiv {\bf U}_0 + {\bm u}$ and ${\bf B} \equiv {\bf B}_0 + {\bf b}$ -- where ${\bm u}$ and ${\bf b}$ are respectively the velocity and magnetic perturbation fields.
The adimensioning of our set of equations is done using the shell thickness $d$ as the reference for length, the Alfv\'en timescale $\tau_{\cal A} = d\,\sqrt{\rho\,\mu}/B_0$ as the reference for time and the Elsasser unit $\sqrt{\rho\,\mu\,\eta\,\Omega}$ as the reference for magnetic field strength -- with $B_0$ the rms value of the background magnetic field's strength, and $\eta$ the magnetic diffusivity of the fluid.
The amplitude of the velocity field is arbitrary.
This system is controlled by three dimensionless parameters, the Lehnert $\lambda$, Lundquist $S$ and magnetic Prandtl $Pm$ numbers, which are respectively defined by
\begin{align}
\label{eq:adim_par}
\lambda = \dfrac{B_0}{\Omega d \sqrt{\rho\,\mu}}\,, \; 
S = \dfrac{d B_0}{\eta \sqrt{\rho\,\mu}}\,, \; 
Pm = \dfrac{\nu}{\eta}\,,
\end{align}
where $\nu$ is the kinematic viscosity of the fluid.
Note that each of these parameters can also be thought of as a timescale ratio, {\it i.e.}
\begin{align}
\label{eq:adim_par_tau}
\lambda = \dfrac{\tau_\Omega}{\tau_{\cal A}}\,, \; 
S = \dfrac{\tau_\eta}{\tau_{\cal A}}\,, \; 
Pm = \dfrac{\tau_\eta}{\tau_\nu}\,,
\end{align}
where $\tau_\eta = d^2/\eta$ is the magnetic diffusive timescale, $\tau_\nu = d^2/\nu$ is the viscous diffusive timescale, and $\tau_\Omega = 1/\Omega$ is the rotation timescale.
For the Earth's core, these parameters are estimated to be approximately $\lambda \sim 10^{-4}$, $S \sim 10^{5}$ and $Pm \sim 10^{-6}$ \cite{de1998viscosity, gillet2010fast, pozzo2014thermal}.
Additionally, an other important dimensionless parameter for the Earth's core dynamics 
is the Ekman number $Ek = \nu / \Omega d^2 = \tau_\Omega / \tau_\nu$, of which we also report the estimate for the Earth's core $Ek \sim 10^{-15}$.

Focusing on the $1^\mathrm{st}$ order dynamics, we neglect the background flow ${\bf U}_0$ -- which is thought to be negligible compared to the Alfv\'en timescales in the Earth's core \cite{gillet2015planetary,baerenzung2018modeling} -- and the Lorentz force $(\nabla \times {\bf B}_0) \times {\bf B}_0$ that could drive a background flow -- a standard choice in such studies ({\it e.g.}, \cite{gillet2011rationale,gerick2021fast}).
The time evolution of the perturbation fields then reads
\begin{align}
\label{eq:momentum_no-T_linearised}
\dfrac{\partial {\bm u}}{\partial t} + \dfrac{2}{\lambda}\,{\bm e}_z \times {\bm u} = - \nabla p + \dfrac{1}{Pm\,\lambda}\,\left[ (\nabla \times {\bf b}) \times {\bf B}_0 + (\nabla \times {\bf B}_0) \times {\bf b} \right] + \dfrac{Pm}{S}\,\nabla^2 {\bm u}\,,
\end{align}
\begin{align}
\label{eq:induction_linearised}
\dfrac{\partial {\bf b}}{\partial t} = \nabla \times ({\bm u} \times {\bf B}_0) + \dfrac{1}{S}\,\nabla^2 {\bf b}\,. 
\end{align}
The perturbations ${\bm u}$ and ${\bf b}$ are triggered by an initial perturbation of the inner core (following \cite{jault2008axial}).

An important quantity in such a setup is the local Alfv\'en speed depending on the strength of the $z$-averaged cylindrical radial component of the background magnetic field $B_{0,s}$, such that
\begin{align}
\label{eq:Va_loc}
V_{\cal A}(s, \phi) \equiv \sqrt{\left< \dfrac{B_{0,s}^2}{\rho\,\mu} \right>}\,,
\end{align}
with $s = r \sin \theta$ the cylindrical radius, and where the angular brackets $\left< x \right>$ in the above expression refer to the axial average of any quantity $x$, such that
\begin{align}
\label{eq:z_average}
\left< x \right> \equiv \dfrac{1}{2h} \displaystyle\int_{-h}^{h}  x\, \mathrm{d}z\,,
\end{align}
with $z = r \cos \theta$ the coordinate along the direction of the rotation axis, and $h \equiv \sqrt{s_o^2 - s^2}$ the half-height of a cylinder aligned with the rotation axis at a cylindrical radius $s$.
$s_o$ is the cylindrical radius of the outer core surface, {\it i.e.} $s_o = r_o$.
An axisymmetric Alfv\'en speed can also be defined as
\begin{align}
\label{eq:Va_ave}
\overline{V}_{\cal A}(s) \equiv \sqrt{\left< \dfrac{\overline{B}_{0,s}^2}{\rho\,\mu} \right>}\,,
\end{align}
where the overbar $\overline{x}$ denotes the azimuthal average of any quantity $x$, {\it i.e.}
\begin{align}
\label{eq:phi_average}
\overline{x} \equiv \dfrac{1}{2 \pi} \displaystyle\int_{0}^{2\pi}  x\, \mathrm{d}\phi\,.
\end{align}

\subsection{Initial conditions}
\label{sec:IC-Impulse}

The system starts at rest with respect to the reference frame of rotation.
The outer core fluid is then accelerated by the Lorentz force after a super-rotating impulse of the inner core.
This impulse forcing follows \cite{jault2008axial,gillet2011rationale} and approaches a Dirac, such that
\begin{align}
\label{eq:impulse_OmegaIC}
\Omega_\mathrm{IC} = \Delta \Omega\,e^{-\left( \dfrac{t}{\tau^*}-3 \right)^2}\,,
\end{align}
where the duration of the forcing is a small fraction of the Aflv\'en time $\tau^* = \tau_\mathrm{pulse}/\tau_{\cal A} = 1.1 \times 10^{-2}$ and remains constant within our set of simulations. 

\subsection{Magnetic base state}
\label{sec:B0}

Again following \cite{jault2008axial,gillet2011rationale}, our static background magnetic field ${\bf B}_0$ is chosen as to respect the insulating condition at the outer boundary, and reads
\begin{align}
\label{eq:B0}
{\bf B}_0 = \nabla \times (\nabla \times W\,{\bf r})\,, \; \mathrm{with} \\
W = \left[ j_1\,(\beta_{11}\,r/r_o) -0.3\,j_1\,(\beta_{12}\,r/r_o) \right] &\,Y_1^0(\theta, \phi) \nonumber \\
    - 0.2\,{j_3}\,(\beta_{31}\,r/r_o) &\,Y_3^0(\theta, \phi) \\
    + 0.3\,{j_3}\,(\beta_{31}\,r/r_o) &\,Y_3^3(\theta, \phi)\,, \nonumber
\end{align}
where $j_n$ is the $n^\mathrm{th}$ order spherical Bessel function of the first kind, and $\beta_{np}$ is the $p^\mathrm{th}$ root of $j_{n-1}(\beta\,r/r_o)$.
$Y_\ell^m(\theta, \phi)$ are the spherical harmonics into which both the velocity and magnetic fields are decomposed with $\ell$ and $m$ respectively the spherical harmonic degree and order of the decomposition.

\begin{figure*}
\centering{
	\includegraphics[width=0.46\linewidth]{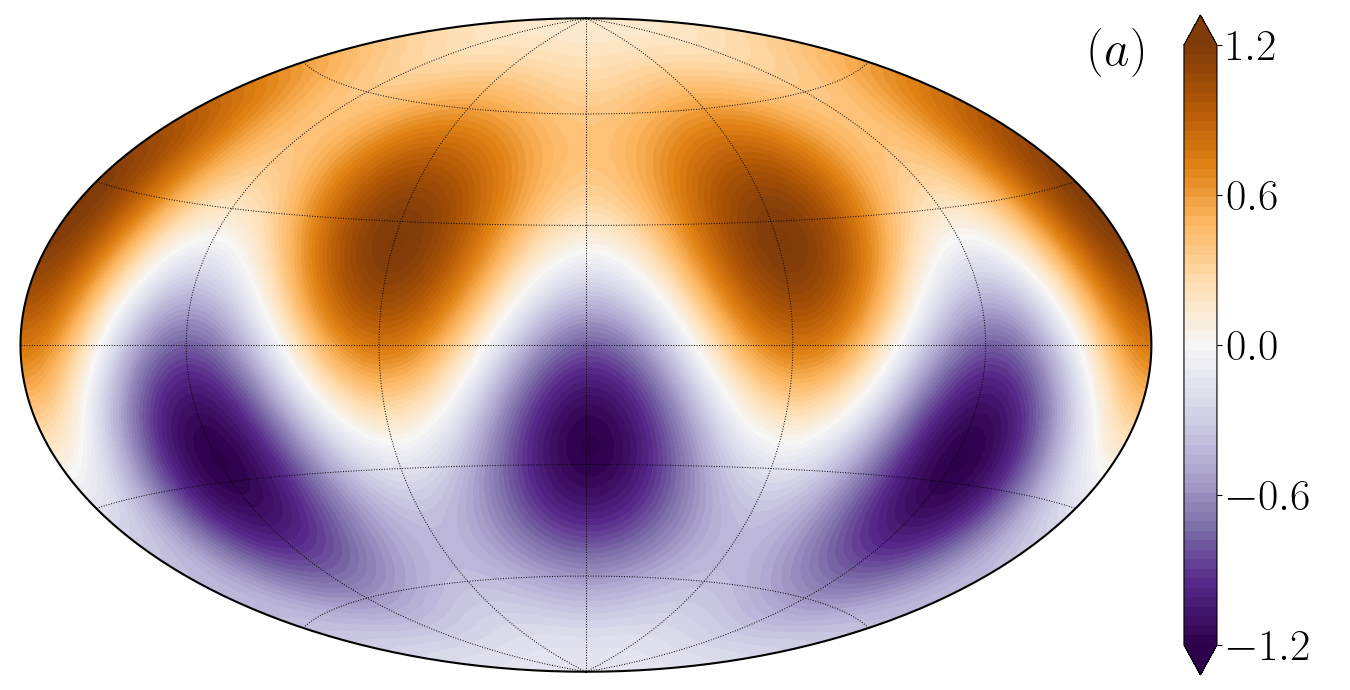}
    \includegraphics[width=0.12\linewidth]{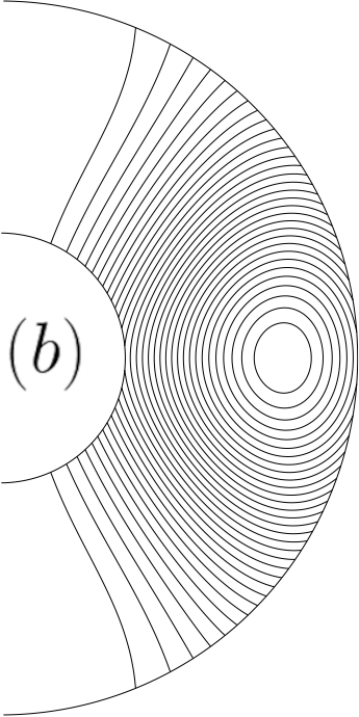}
    \includegraphics[width=0.38\linewidth]{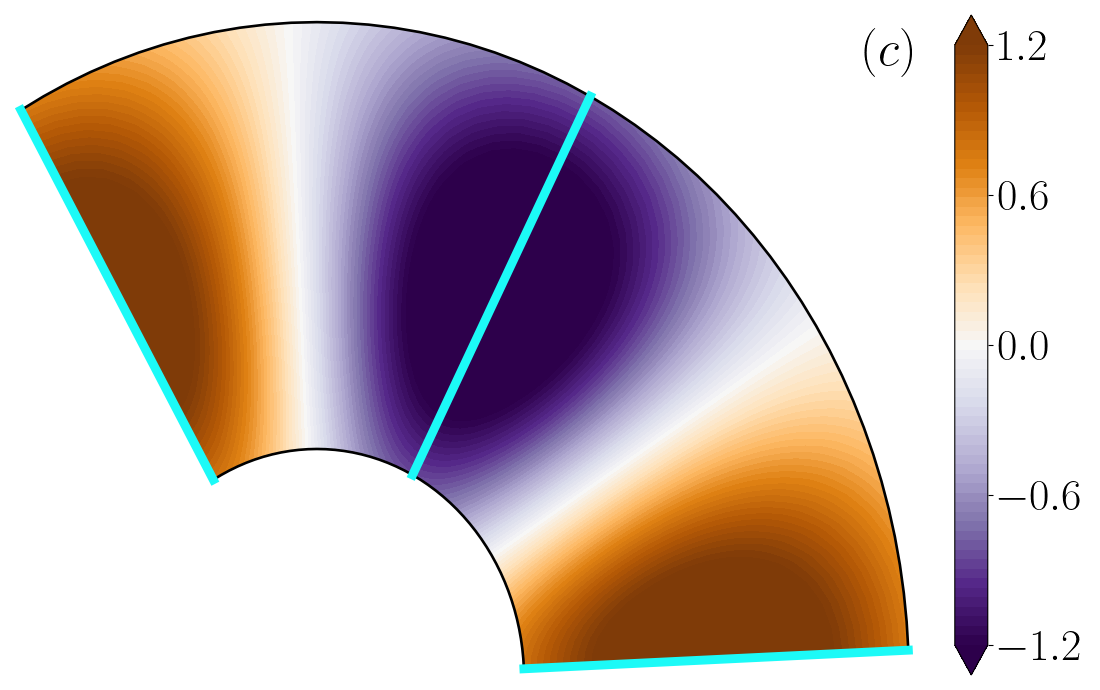}
}
	\caption{
	Radial component of the background magnetic field at the CMB $B_{0,r}\vert_{r = r_o}$ (a),
    fieldlines of the axisymmetric poloidal background magnetic field (b),
    and radial component of the background magnetic field in the equatorial plane $B_{0,r}\vert_{\theta = \pi / 2}$ (c).
    The trajectories along fast longitudes (in light-blue) are additionally shown in the equatorial plane. 
	}
	\label{fig:B_background}
\end{figure*}

The radial component of ${\bf B}_0$ at the core mantle boundary, the fieldlines of the axisymmetric poloidal magnetic field and an equatorial section of the radial component of this background magnetic field are displayed in Figure~\ref{fig:B_background}.
The Alfv\'en speed $V_{\cal A}$ depending on the strength of the background magnetic field (see eq.~\ref{eq:Va_loc}), trajectories along fast longitudes -- defined as where $||{\bf B}_0||$ is the largest -- (in light-blue) are displayed for $B_{0,r}$ in the equatorial plane (Fig.~\ref{fig:B_background}~c). 
This base state was first designed to aim at modelling torsional waves in the Earth's outer core \cite{jault2008axial}.
But the main characteristics that interest us in this study are: ({\it i}) it is non-axisymmetric, ({\it ii}) it has a non-zero $B_{0,r}^2$ component at the equator, and ({\it iii}) it matches a potential outer boundary condition.

\subsection{Numerical implementation}
\label{sec:Numerics}

Our numerical implementation is derived from \cite{aubert2013bottom, aubert2017spherical} and relies on a spectral spherical harmonic decomposition of both the velocity and magnetic fields up to a maximum degree and order $\ell_\mathrm{max} = m_\mathrm{max}$ in the horizontal direction while a discretisation using a second-order finite-difference scheme up to a maximum of $N_r$ grid points is used in the radial direction.
To handle the spherical harmonic transforms, we use the open-source \texttt{SHTns}\footnote{\url{https://bitbucket.org/nschaeff/shtns}} library \cite{schaeffer2013efficient}.
Parallelisation of the code relies on the Message Passing Interface ({\tt MPI}) library.
A second-order, semi-implicit scheme is used to time-step the equations of the system.

We follow \cite{aubert2017spherical,aubert2019approaching} and we approximate our solutions using an hyperdiffusivity approach applied to the small length-scales of the velocity field -- the magnetic field remaining fully resolved.
The numerical implementation of this hyperdiffusivity is derived from \cite{nataf2015turbulence} and takes the form
\begin{align}
\label{eq:hdif-vel}
\nu_\mathrm{eff} =  \nu\,q_H^{\ell-\ell_H}\, \; \mathrm{for }\, \; \ell \geq \ell_H\,,
\end{align}
with $\ell_H$ the cut-off degree above which the hyperdiffusion smoothly increases and affects the small length-scales of the solution at a rate controlled by the parameter $q_H$.
Here, we adopt values for the hyperdiffusion parameters that have been used in previous studies showing a satisfying convergence of the magnetic and kinetic energy spectra ({\it e.g.}, \cite{aubert2019approaching}).
Note that in order to test the sensitivity of our results on the hyperdiffusion parameters we have computed several cases varying the values of $\ell_H$ and $q_H$ without observing major changes in the average properties.
More details about the hyperdiffusion approach and its geophysical justifications can be found in \cite{aubert2017spherical}.
Values of $\ell_\mathrm{max}$, $N_r$, $\ell_H$ and $q_H$ used in our models are reported in Table~\ref{tab:run_list}.

\subsection{Diagnostics}
\label{sec:diagnostics}

We introduce in this section notations for various diagnostics.
First, we define the dimensionless kinetic energy as
\begin{linenomath}
\begin{align}
\label{eq:E_kin}
E_\mathrm{kin} = \dfrac{1}{2} {\bm u}^2\,, 
\end{align}
\end{linenomath}
as well as the dimensionless non-axisymmetric (non-zonal) and axisymmetric (zonal) kinetic energies, respectively
\begin{align}
\label{eq:E_kin_fluct-zon}
\tilde{E}_\mathrm{kin} = \dfrac{1}{2} \tilde{\bm u}^2\,, \; 
\overline{E}_\mathrm{kin} = \dfrac{1}{2} \overline{u}_\phi^2\,,
\end{align}
where the tilde $\tilde{x}$ stands for the non-axisymmetric component of any quantity $x$, such that
\begin{align}
\label{eq:field_fluct}
\tilde{x} \equiv x - \overline{x}\,.
\end{align}
We similarly define the dimensionless magnetic energies as
\begin{align}
\label{eq:E_mag}
E_\mathrm{mag} = \dfrac{1}{2} \dfrac{1}{Pm\,\lambda} {\bf b}^2\,, \; 
\tilde{E}_\mathrm{mag} = \dfrac{1}{2} \dfrac{1}{Pm\,\lambda} {\bf \tilde{b}}^2\,, \; 
\overline{E}_\mathrm{mag} = \dfrac{1}{2} \dfrac{1}{Pm\,\lambda} \overline{b}_\phi^2\,.
\end{align}

\section{Taxonomy of waves}
\label{sec:Waves-types}

We summarise here and in Table \ref{tab:wave_types} the types of waves that can propagate in the configuration defined in the previous section (see \cite{gillet2022dynamical} for a more complete overview).
An equilibrium between inertia and the Lorentz perturbation (QG-IM) can produce either axisymmetric or non-axisymmmetric waves.
The axisymmetric wave is the torsional Alfv\'en wave \cite{braginsky1970torsional} which moves along co-axial cylinders parallel to the tangent cylinder, has an energetic equipartition $E_\mathrm{kin} = E_\mathrm{mag}$, is non-dispersive, and evolves on annual to decadal timescales comparable to $\tau_{\cal A}$.
The QG-Alfv\'en waves have the same characteristics as the torsional wave, {\it i.e.} have an energetic equipartition $E_\mathrm{kin} = E_\mathrm{mag}$, are non-dispersive, and evolve on timescales comparable to $\tau_{\cal A}$, but are non-axisymmetric.
An equilibrium between inertia and the Coriolis perturbation (QG-IC) yields the slow inertial waves or 'slow' quasi-geostrophic Rossby waves \cite{zhang1993equatorially,zhang2001inertial} (hereafter simply Rossby waves) which are non-axisymmetric, dispersive, have $E_\mathrm{kin} \gg E_\mathrm{mag}$, and evolve on sub-annual timescales faster than $\tau_{\cal A}$.
And a QG-MC equilibrium can occur between the Lorentz and Coriolis perturbations, producing the magneto-Coriolis waves \cite{hide1966free,malkus1967hydromagnetic,gerick2021fast} which are non-axisymmetric, dispersive, have $E_\mathrm{kin} \ll E_\mathrm{mag}$, and evolve on multi-annual timescales comparable to $\tau_{\cal A}$ up to millennial timescales much slower than $\tau_{\cal A}$.

Finally, note that all the aforementioned waves can be modified by dissipative effects -- affecting their decay or their reflection at the boundaries -- but as this study focuses on the transient dynamics, the dissipation of these waves will not be addressed in details.

\begin{table}
\caption{Summary of the various types of waves expected in our configuration with some of their properties like their corresponding force balance, their main direction of propagation, their period relative to the Alfv\'en timescale $\tau_{\cal A}$ or the ratio of their kinetic to magnetic energies $E_\mathrm{kin}/E_\mathrm{mag}$.}
\label{tab:wave_types}
\centering
\begin{tabular*}{0.995\columnwidth}{@{}l@{\extracolsep{\fill}}cccccc}
\hline
 Wave Type            & Balance  & Axisymmetric  & Dispersive    & Propagation  & Period                   & $E_\mathrm{kin}/E_\mathrm{mag}$ \\
\hline
 Slow Rossby          & QG-IC    & $\times$	    & $\checkmark$  & $\phi$       & $< \tau_{\cal A}$      & $ \gg 1$ \\
 QG-Magneto-Coriolis  & QG-MC    & $\times$      & $\checkmark$	& $\phi$       & $\gtrsim \tau_{\cal A}$  &  $\ll 1 \; \mathrm{to} \; 1$ \\
 Torsional Aflv\'en	 & QG-IM    & $\checkmark$	& $\times$	    & $s$          & $\sim \tau_{\cal A}$     &  $1$ \\
 QG-Alfv\'en            & QG-IM    & $\times$	    & $\times$	    & $s$          & $\sim \tau_{\cal A}$     &  $1$ \\
\hline
\end{tabular*}
\end{table}

\section{Results}
\label{sec:Results}

We have computed 23 simulations, between $Ek= 3 \times 10^{-7}$ and $Ek= 3 \times 10^{-10}$ covering a range of Lundquist numbers $500 \leq S \leq 8000$ and Lehnert numbers $2.2 \times 10^{-4} \leq \lambda \leq 3.5 \times 10^{-3}$ (see \S~\ref{sec:MHD-lin_equations} for comparison with the Earth's core values).
Parameter values are summarised in Table~\ref{tab:run_list} in the Appendix section.

\subsection{Temporal evolution}
\label{sec:Time_evo}

\begin{figure*}
\centering{
	\includegraphics[width=.99\linewidth]{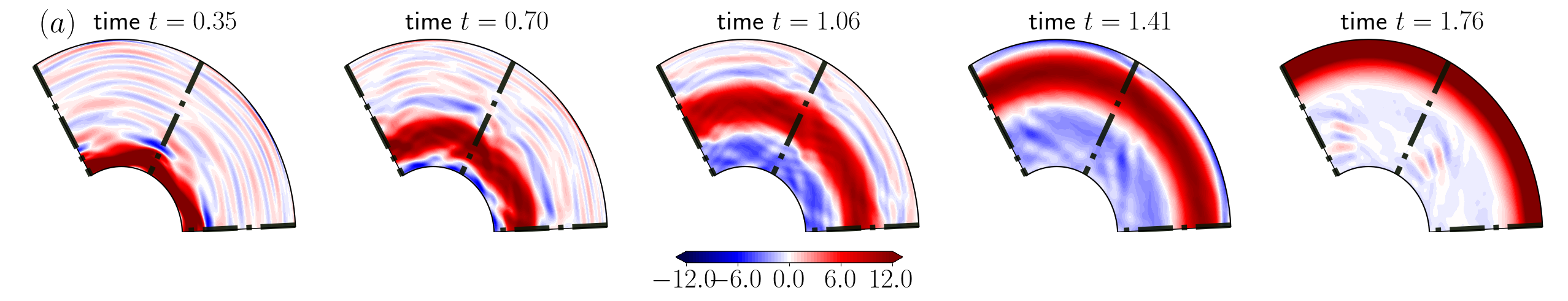}}
\centering{
	\includegraphics[width=.99\linewidth]{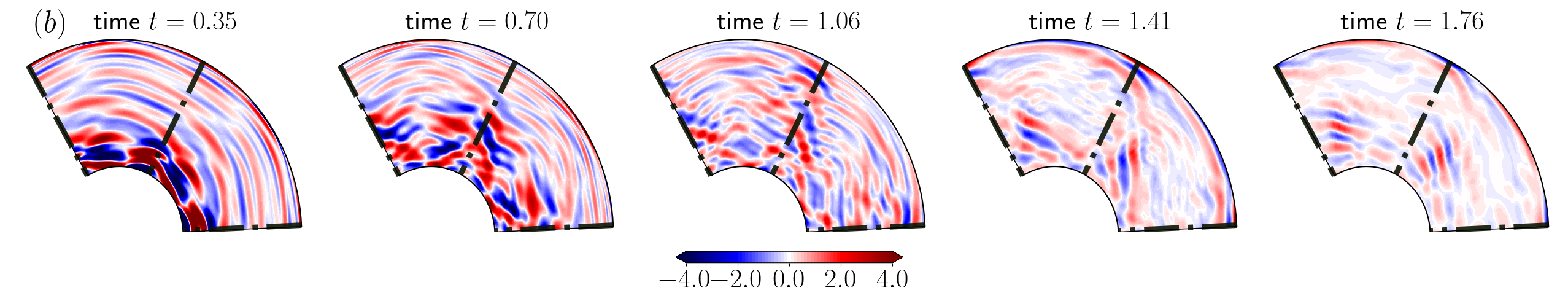}}
\centering{
	\includegraphics[width=.99\linewidth]{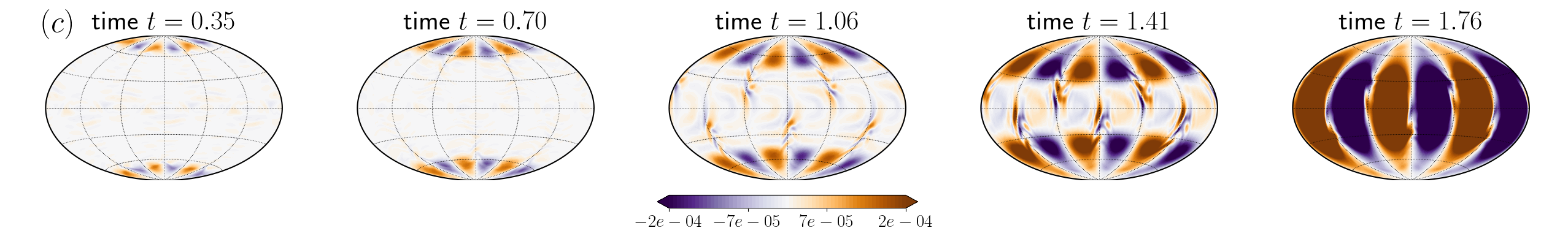}}
	\caption{
	Series of snapshots of the total azimuthal velocity field in the equatorial plane $u_\phi(\theta = \mathrm{equator})$ (a), of the non-axisymmetric azimuthal velocity field in the equatorial plane $\tilde{u}_\phi(\theta = \mathrm{equator})$ (b), and of the radial perturbation magnetic field at the outer boundary $b_r(r = r_o)$ (c) for a case at $Ek= 10^{-7}$, $Pm=0.144$ and $S = 1596$.
    The trajectories along fast longitudes (dotted black lines) are additionally shown in the equatorial plane plots (panels a-b).
    Times are given in terms of the Alfv\'en timescale $\tau_{\cal A}$.
	}
	\label{fig:Maps_t_evolution}
\end{figure*}

In Figure~\ref{fig:Maps_t_evolution}~(a), we can observe that a thick outward propagating axisymmetric wave, crossing the shell in about $t \sim 1.76\,\tau_{\cal A}$, appears to be the main feature in the total flow $u_\phi$.

Together with the axisymmetric wave, a series of non-axisymmetric waves localised in $s$ and $\phi$ are emitted after the initial impulse and appear to propagate outwards faster than the axisymmetric wave (Fig.~\ref{fig:Maps_t_evolution}~b).
At the beginning, these waves keep a comparable radial and azimuthal wavelengths but they become broader in the $\phi$-direction and thinner in the $s$-direction as they approach the outer boundary.
Fainter azimuthally elongated waves also appear at the outer boundary shortly after the start of the simulation, and propagate inwards.

Looking at $b_r(r_o)$ (Fig.~\ref{fig:Maps_t_evolution}~c) just after the start of the simulation, we can observe a signal close to the poles with a clear $m=3$ azimuthal symmetry.
This signal slowly progresses towards the equator and after $t = 1.06\,\tau_{\cal A}$, we start observing signals with shorter spatial wavelengths in the equatorial region. The equatorially localised wave front clearly displays a westward drift between $t = 1.06\,\tau_{\cal A}$ and $t = 1.41\,\tau_{\cal A}$.
At the end (at $t = 1.76\,\tau_{\cal A}$), only the strongest and slowest signal remains, similarly to what is observed for the total azimuthal flow.

Given the typology of the waves that are expected in our setup (see \S~\ref{sec:Waves-types}), we can deduce that the strongest and slowest signal observed in $u_\phi$ and $b_r(r_o)$ (but not in $\tilde{u}_\phi$) is the torsional Alfv\'en wave emitted at the inner core and slowly moving towards the equator.
The chosen magnetic field background and initial impulse -- a super-rotation of the inner core -- mostly promote, by construction, torsional oscillations that are carried by the azimuthal component of the velocity field and are also clearly visible in the radial component of the magnetic field.
We can also infer that the faster signals with shorter spatial wavelengths, visible both in $\tilde{u}_\phi$ and $b_r$, are the QG-Alfv\'en and QG-Magneto-Coriolis waves.
The signals closer to the equator, extended in $\phi$ and displaying a westward drift should be the QG-MC waves. 
And we can deduce that the fastest and fainter signal at the start of the simulation are the Rossby waves since they predominantly 
carry a flow signature and the magnetic signature is subdominant.

Interested in the identified QG-MC wave front -- visible near the core surface and in the equatorial region after $t = 1.06\,\tau_{\cal A}$ in Fig.~\ref{fig:Maps_t_evolution}~(b,c) --, we investigate some of its properties.
Based on observations of the temporal evolution of the radial magnetic field in the equatorial plane at the core surface $b_r\vert_{\theta = \pi/2} (r = r_o, \phi, t)$ (not shown), we find that the westward drift of the QG-MC waves has a phase speed of $V_\phi \sim 1100\,km/y$.
Similarly, observing $\left< \tilde{u}_\phi \right> (s, \phi = 0, t)$ (not shown), we find that the QG-MC waves near the CMB propagate outward with a speed of $V_s \sim 280\,km/y$.
Both estimates have been redimensionalised for the Earth's core.

\subsection{Torsional wave}
\label{sec:TO}

The complexity of our system identified in the previous section is better approached by examining flows instead of the magnetic field (because they are QG) and in time-cylindrical radius diagrams.
Note that, in these diagrams, with our choice of adimensionalisation $s_i \simeq 0.54$ and $s_o \simeq 1.54$.

\subsubsection{Columnar zonal flow}
\label{sec:TO-diag}

\begin{figure*}
\centering{
	\includegraphics[width=0.49\linewidth]{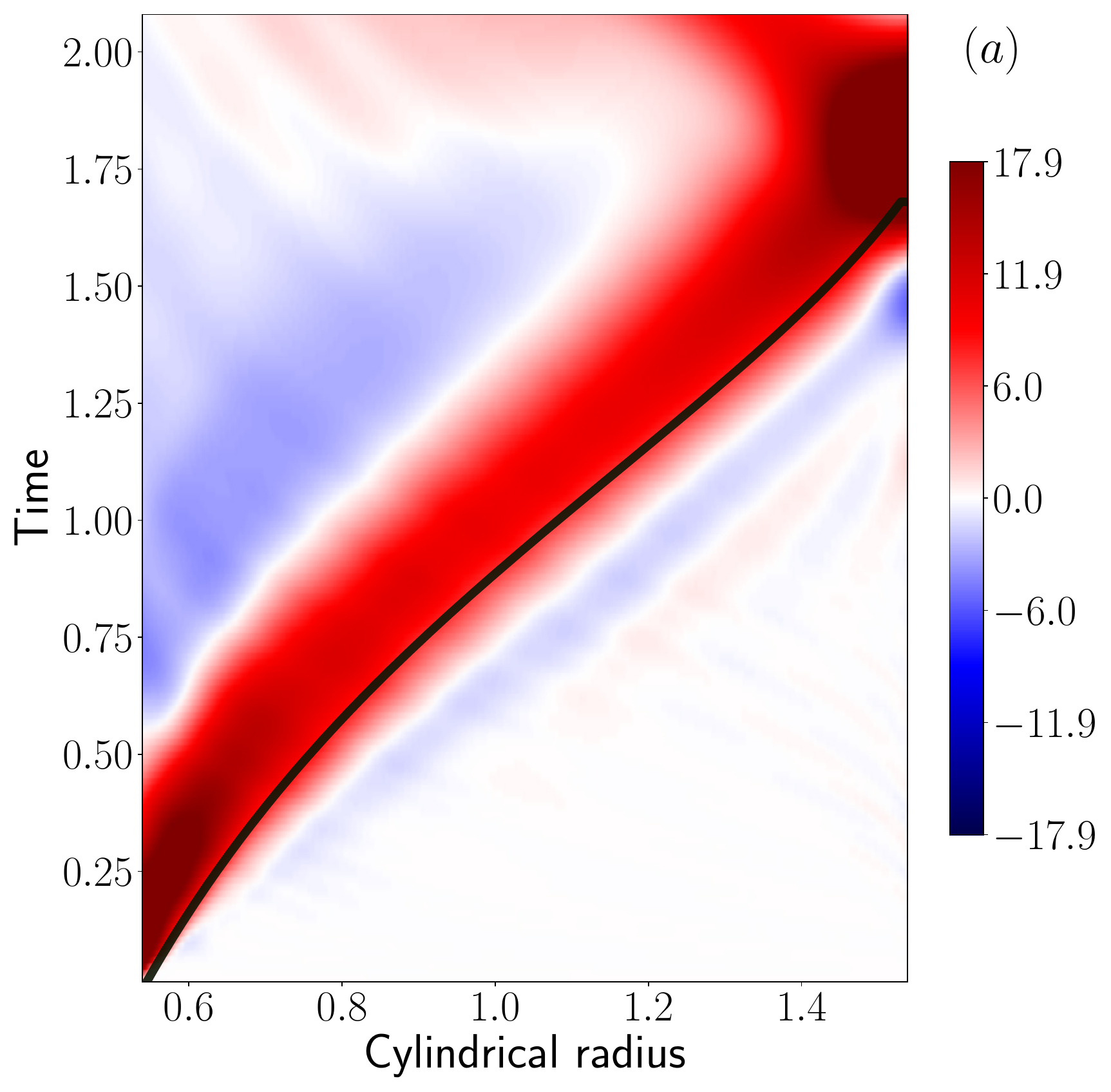}
    \includegraphics[width=0.49\linewidth]{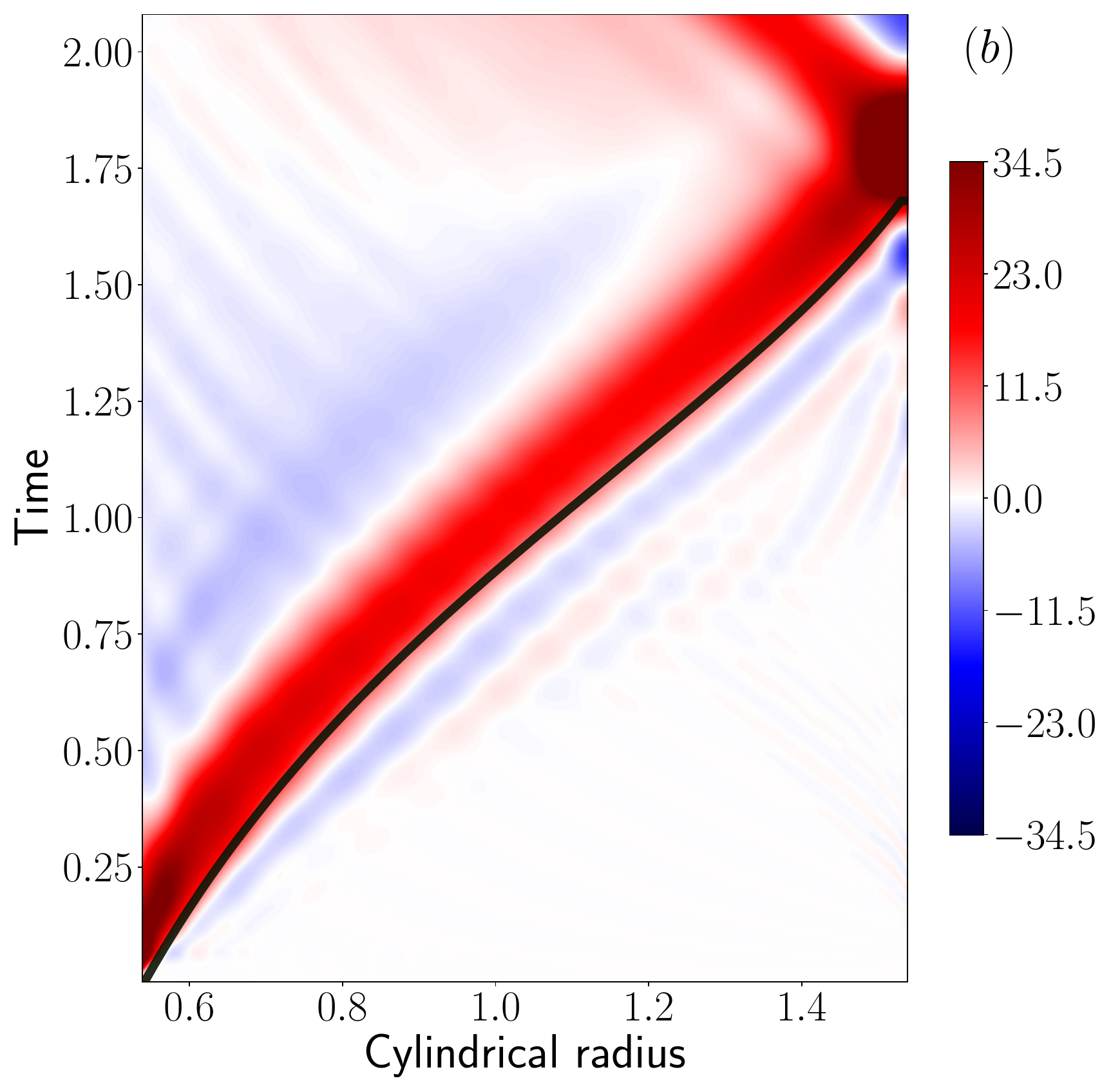}
}
	\caption{
	Temporal evolution of the axially averaged zonal flow $\left< \overline{u}_\phi \right>$ for a case at $Ek= 10^{-7}$, $Pm=0.144$ and $S = 1596$ (a) and a case at $Ek= 3 \times 10^{-10}$, $Pm=7.9 \times 10^{-3}$ and $S = 6825$ (b).
    The trajectories for the $\phi$-averaged Alfv\'en velocity $\overline{V}_{\cal A}(s)$ (in black) is also shown -- it has been obtained by solving for the eikonal equation for a varying Alfv\'en velocity.
	}
	\label{fig:Torsional_oscillation}
\end{figure*}

In Figure~\ref{fig:Torsional_oscillation}, for both cases, we can observe a single wave departing from the inner core boundary and reaching the outer boundary in less than two Alfv\'en times (around $t \sim 1.6\,\tau_{\cal A}$).
In both cases, the signal appears to reflect at the boundary.
Tests using a longer time window reveal that the signal travels back and forth for at least several Alfv\'en times while slowly dissipating (not shown).
We can also see that the wave propagation is compatible with a wave packet travelling at the $\phi$-averaged Alfv\'en velocity $\overline{V}_{\cal A}$ (black trace).
This signal, triggered by the impulse of the inner core and mostly carried by the zonal flow, is the torsional Alfv\'en wave and has already been identified in Fig.~\ref{fig:Maps_t_evolution}~(a).
We can see that the speed of the torsional wave is independent of the parameters.
Indeed, the time arrival of the torsional wave, once rescaled with the Alfv\'en timescale, is the same in all our cases and the wave always reaches the outer boundary around the same time $t_\mathrm{TW-arrival} \simeq 1.65 \tau_{\cal A}$.

\subsubsection{Scaling law for torsional wave thickness}
\label{sec:TO-scaling_law}

\begin{figure*}
\centering{
	\includegraphics[width=0.99\linewidth]{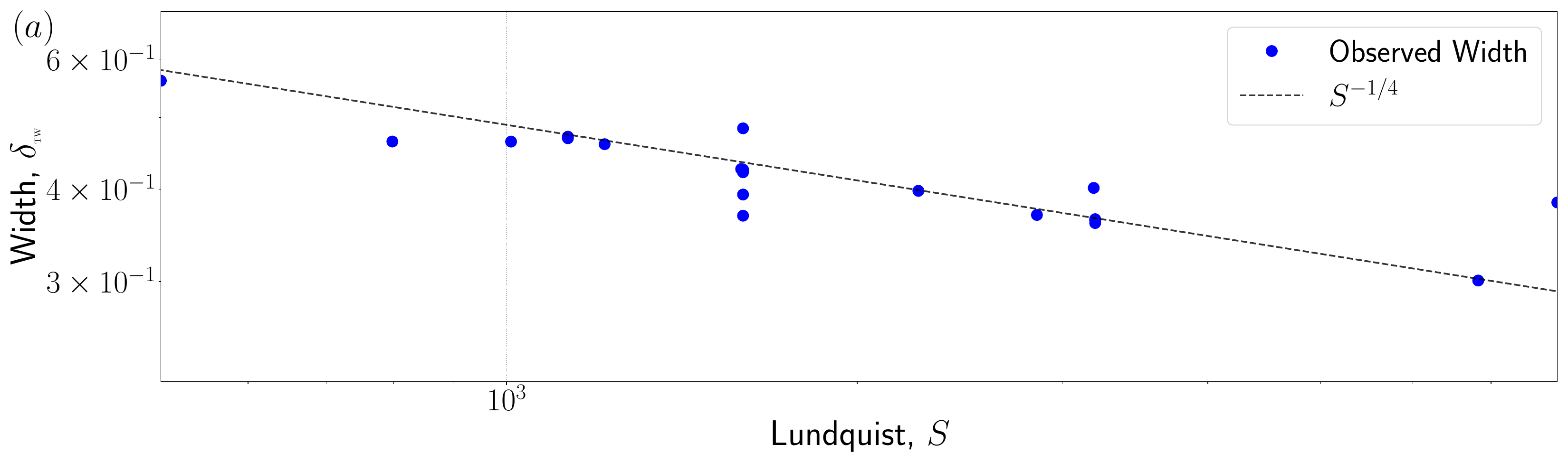}
}
\centering{
	\includegraphics[width=0.99\linewidth]{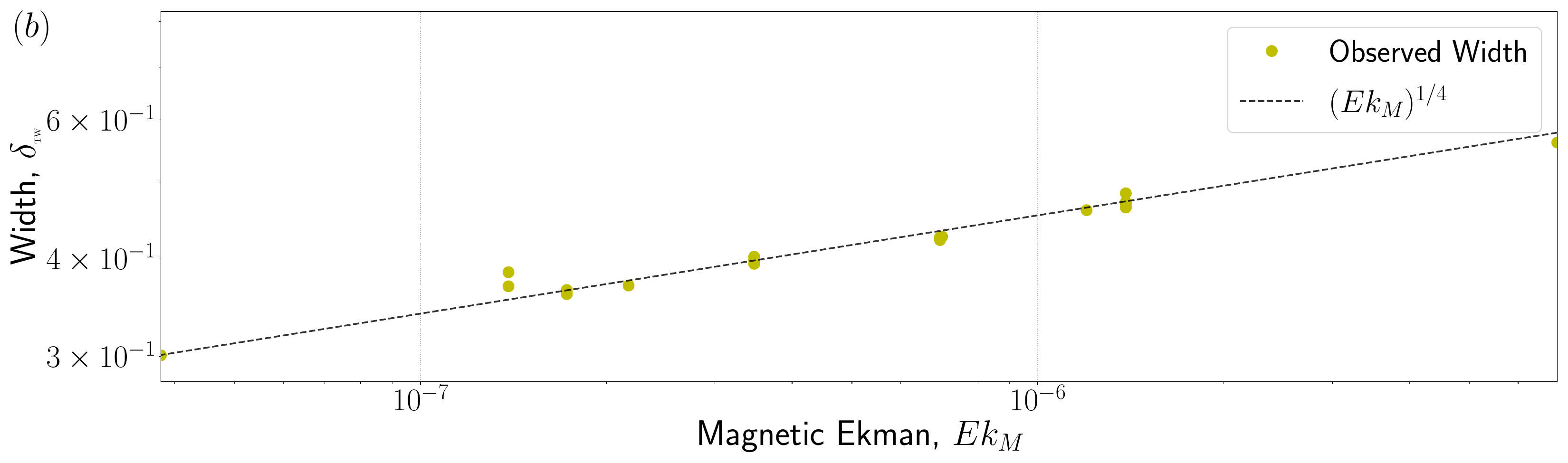}
}
	\caption{
	Scaling of the width of the torsional wave with respect to the Lundquist number $S$ (a), and with respect to the magnetic Ekman number $Ek_M$ (b) for all the cases computed in this study.
    Dotted lines with the expected slope derived in \cite{jault2008axial} are also shown as a comparison.
	}
	\label{fig:Scaling-law_width-vs-S}
\end{figure*}

The width of the torsional wave is expected to depend on the parameters.
\cite{jault2008axial} reported that the width of the torsional wave should follow a power law of the form $S^{-1/4}$. 
This appears to be the case in Fig.~\ref{fig:Torsional_oscillation} as we can observe that the width of the signal is divided by $\sim 1.5$ between the 2 cases (compatible with $S^{-1/4}$). 
The scaling for the thickness of the torsional wave involves an equatorial singularity and the shear layer at the inner boundary \cite{jault2008axial}, and in this paper, the author proposed several empirical and competing scaling laws which read
\begin{align}
\label{eq:Scaling-law_width-vs-S-and-EkM}
\delta_\mathrm{TW} \sim S^{-1/4}\,, \; \mathrm{or}\, \; \delta_\mathrm{TW} \sim (Ek_M)^{1/4}\,,
\end{align}
where $Ek_M \equiv Ek / Pm = \lambda / S$ is the magnetic Ekman number.
We evaluate in Figure~\ref{fig:Scaling-law_width-vs-S} the proposed scaling laws for the characteristic width of the torsional Alfv\'en wave $\delta_\mathrm{TW}$ which is arbitrarily defined at a given time (about $t \sim 1 \tau_{\cal A}$) as the distance along the $s$-direction between the two sides of the torsional wave where its angular velocity reaches zero.
The discrimination between the two proposals, and the influence of the magnetic Prandtl number, had already been questioned by \cite{jault2008axial} and the author favored then the Lundquist number scaling law.
We can see however that the scaling law as a function of the magnetic Ekman number clearly collapses all the considered points better. 

This indicates that the width of the torsional wave depends on rotation and magnetic diffusivity and that its variations are better accounted for by the magnetic Ekman number $Ek_M$, following $\delta_\mathrm{TW} \sim (Ek_M)^{1/4}$.
As $Ek_M = \tau_\Omega / \tau_\eta$, we can understand the thickness of the torsional waves as mostly invariant with respect to $\tau_{\cal A}$ and controlled by how many rotations can fit in a diffusion time.

\subsection{Columnar force balance}
\label{sec:col_bal}

\begin{figure*}
\centering{
    \includegraphics[width=0.31\linewidth]{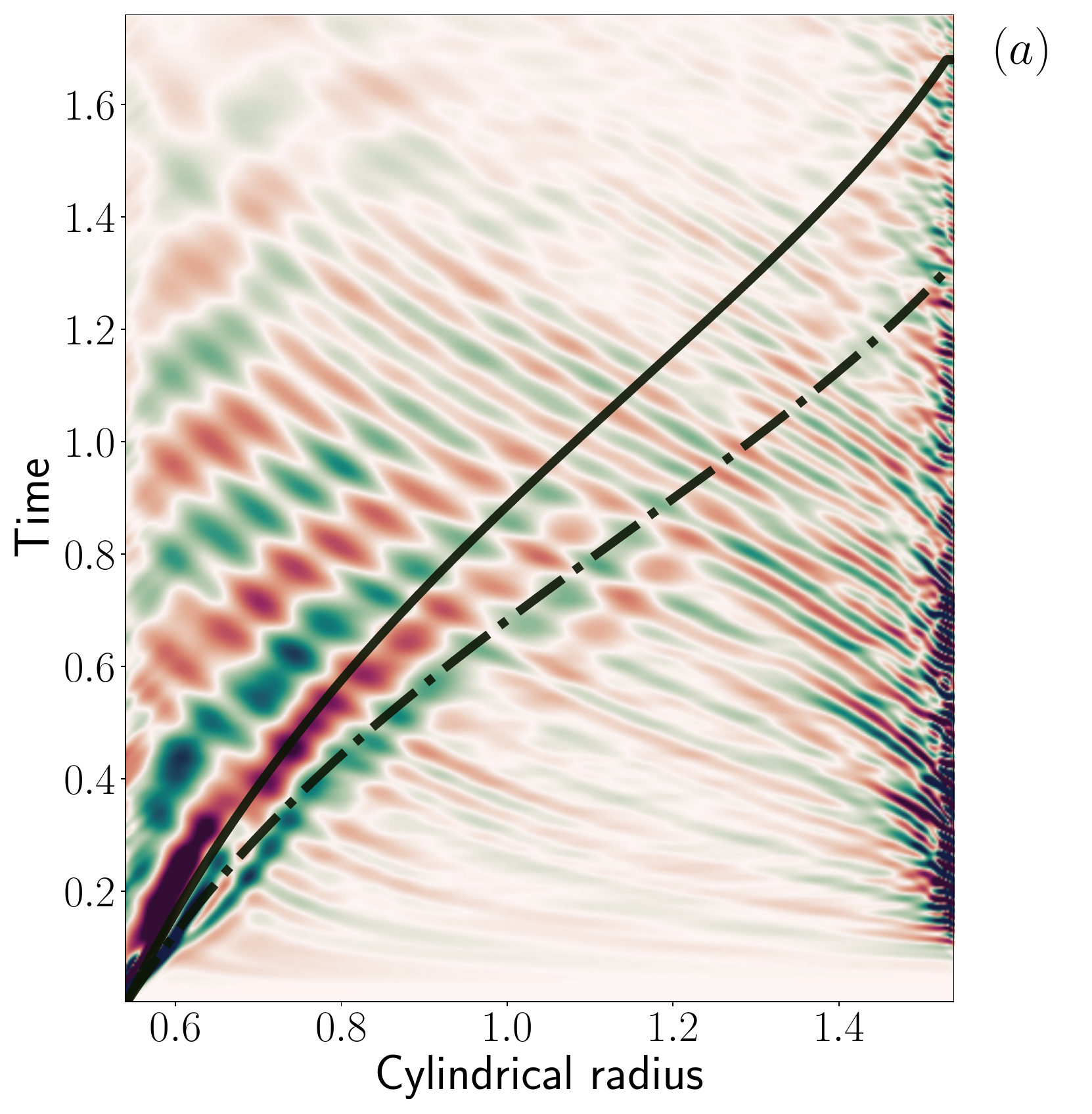}
    \includegraphics[width=0.31\linewidth]{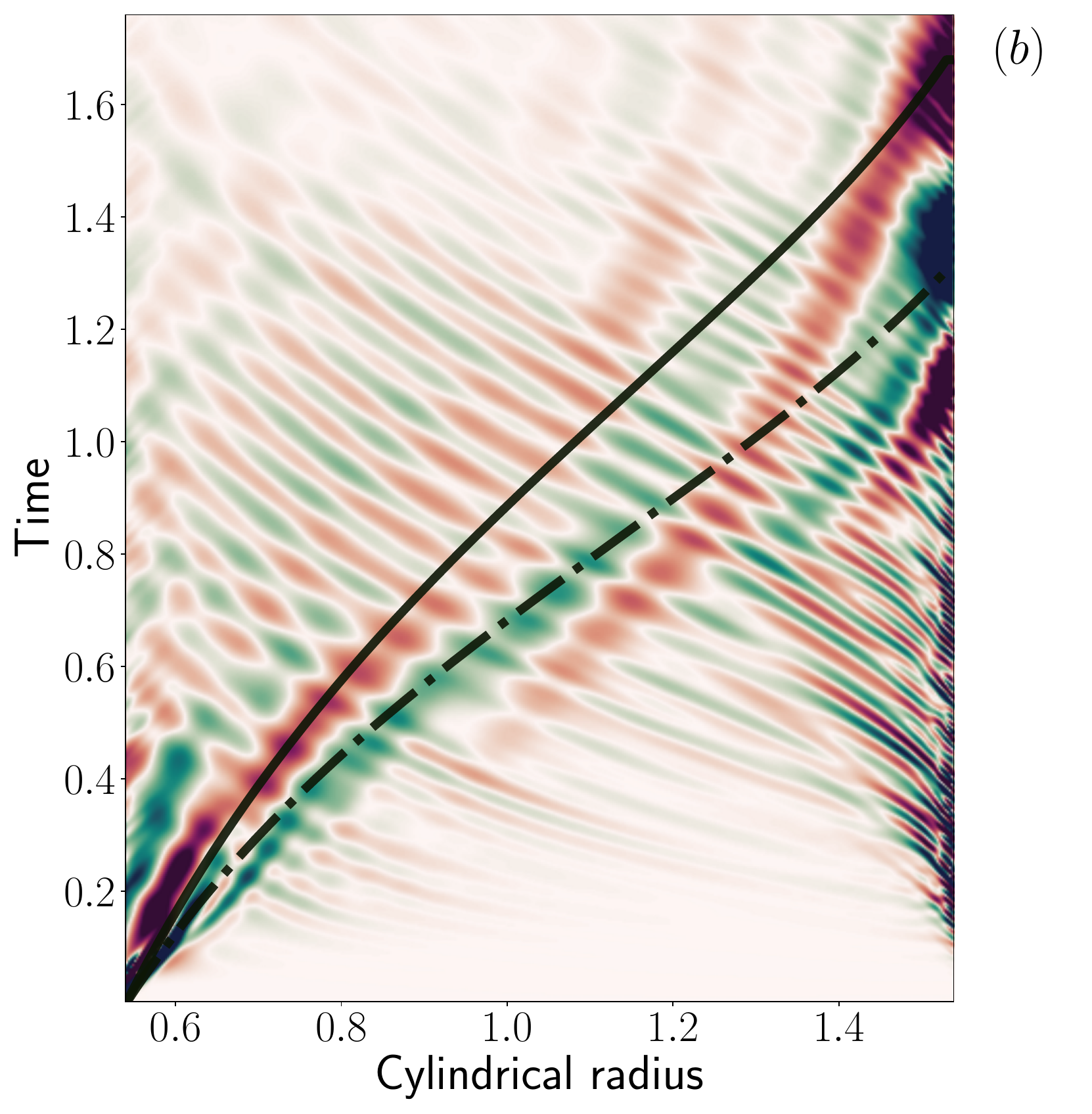}
	\includegraphics[width=0.33\linewidth]{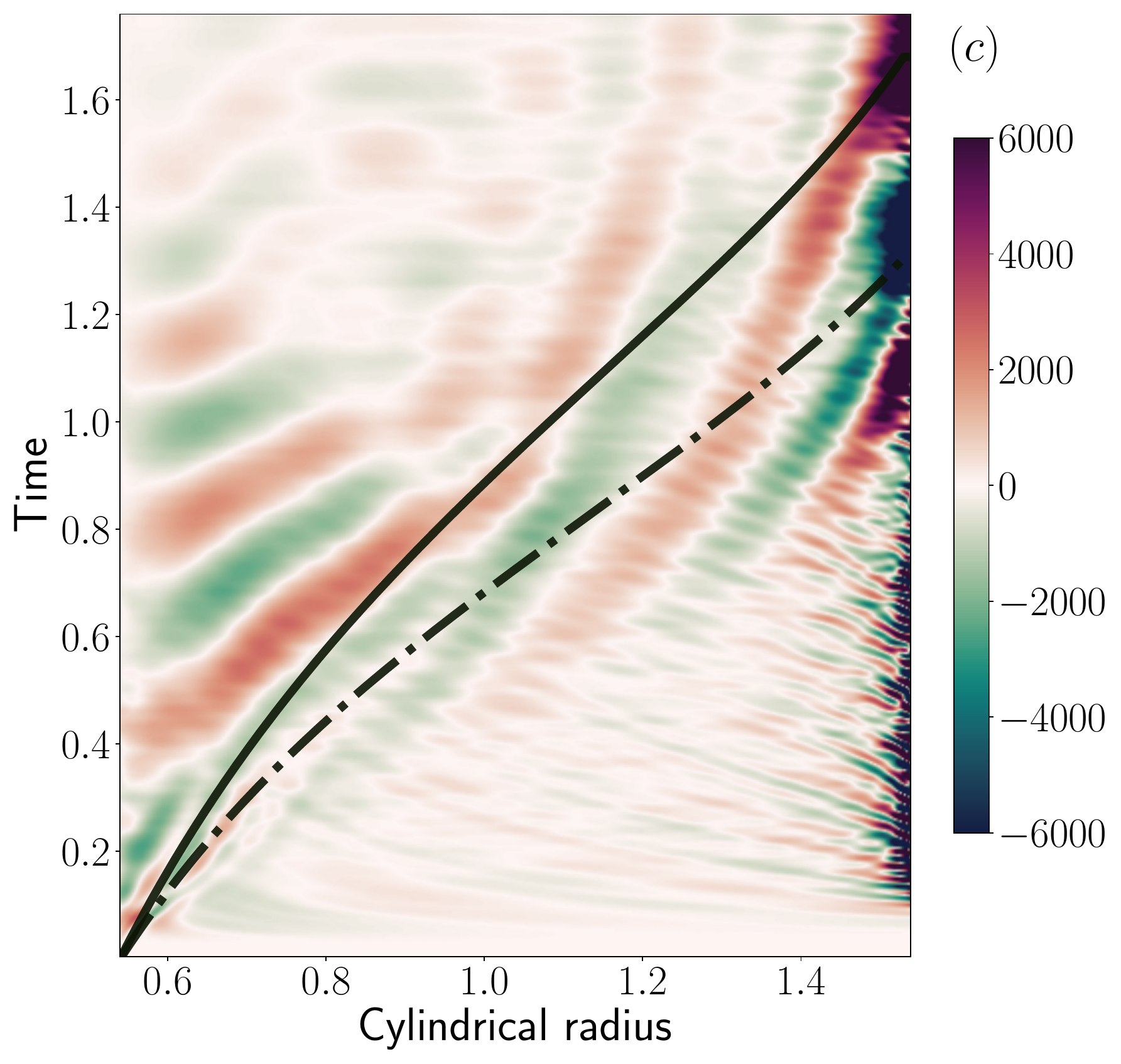}
}
\centering{
    \includegraphics[width=0.31\linewidth]{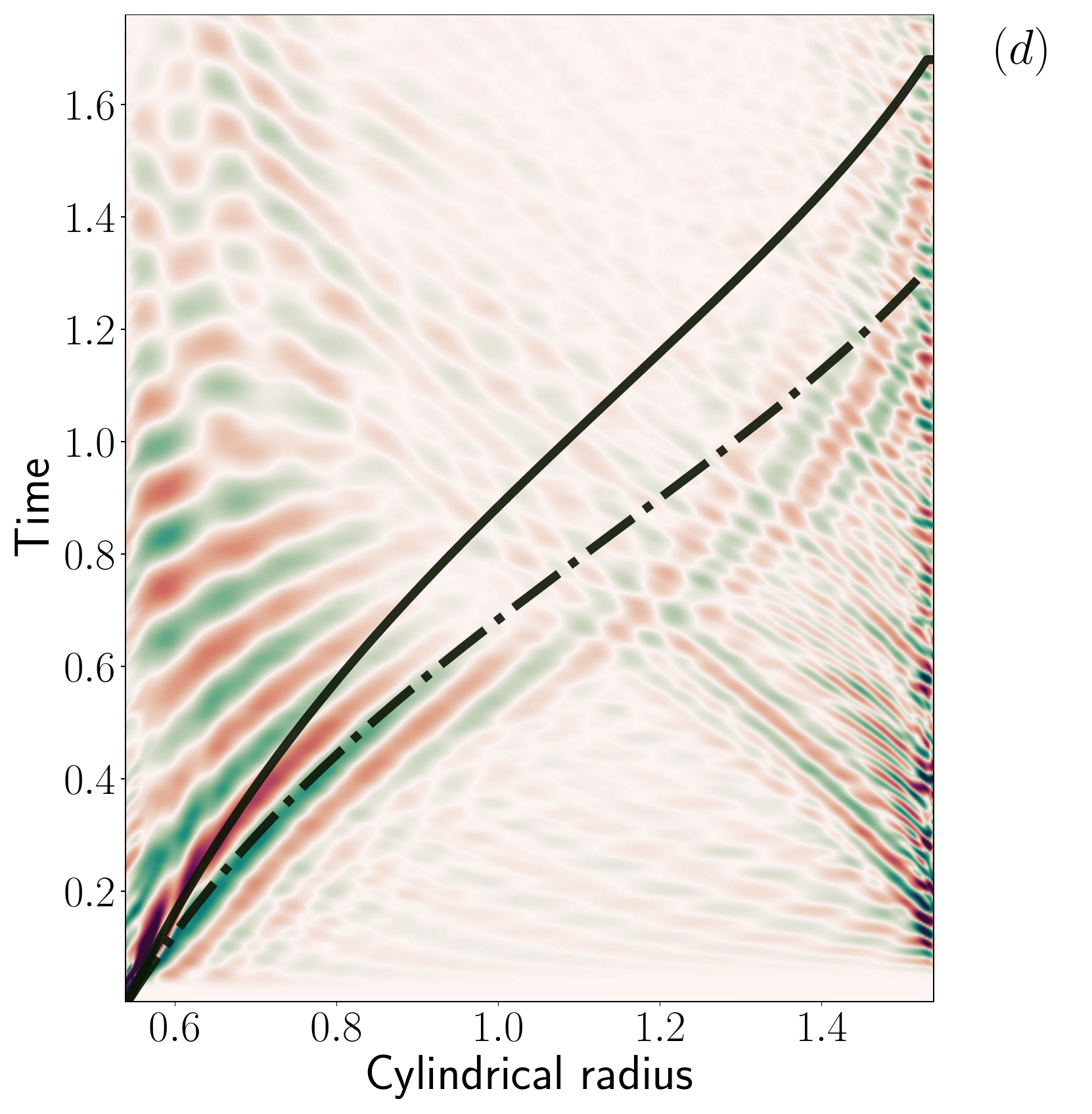}
    \includegraphics[width=0.31\linewidth]{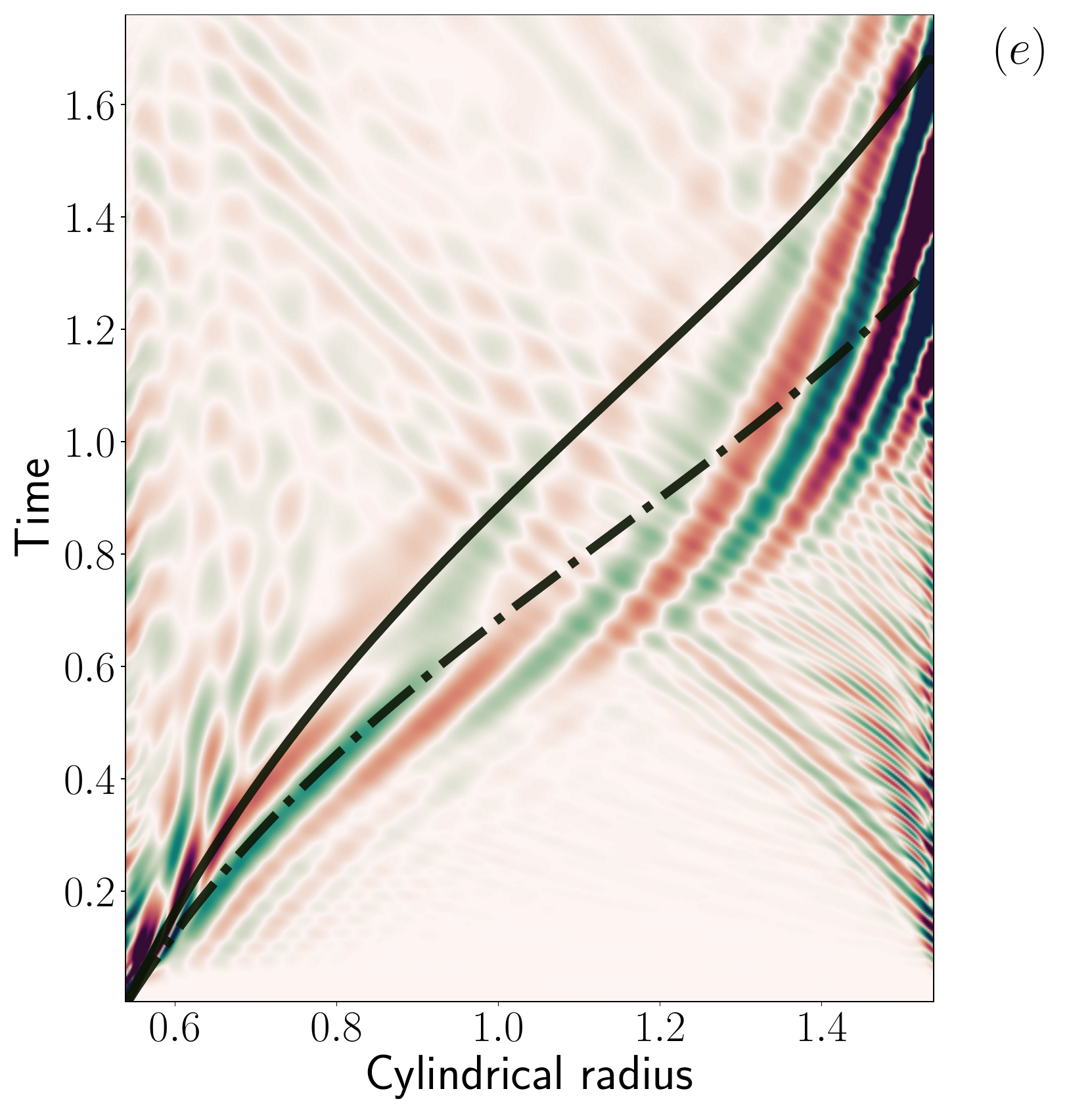}
	\includegraphics[width=0.33\linewidth]{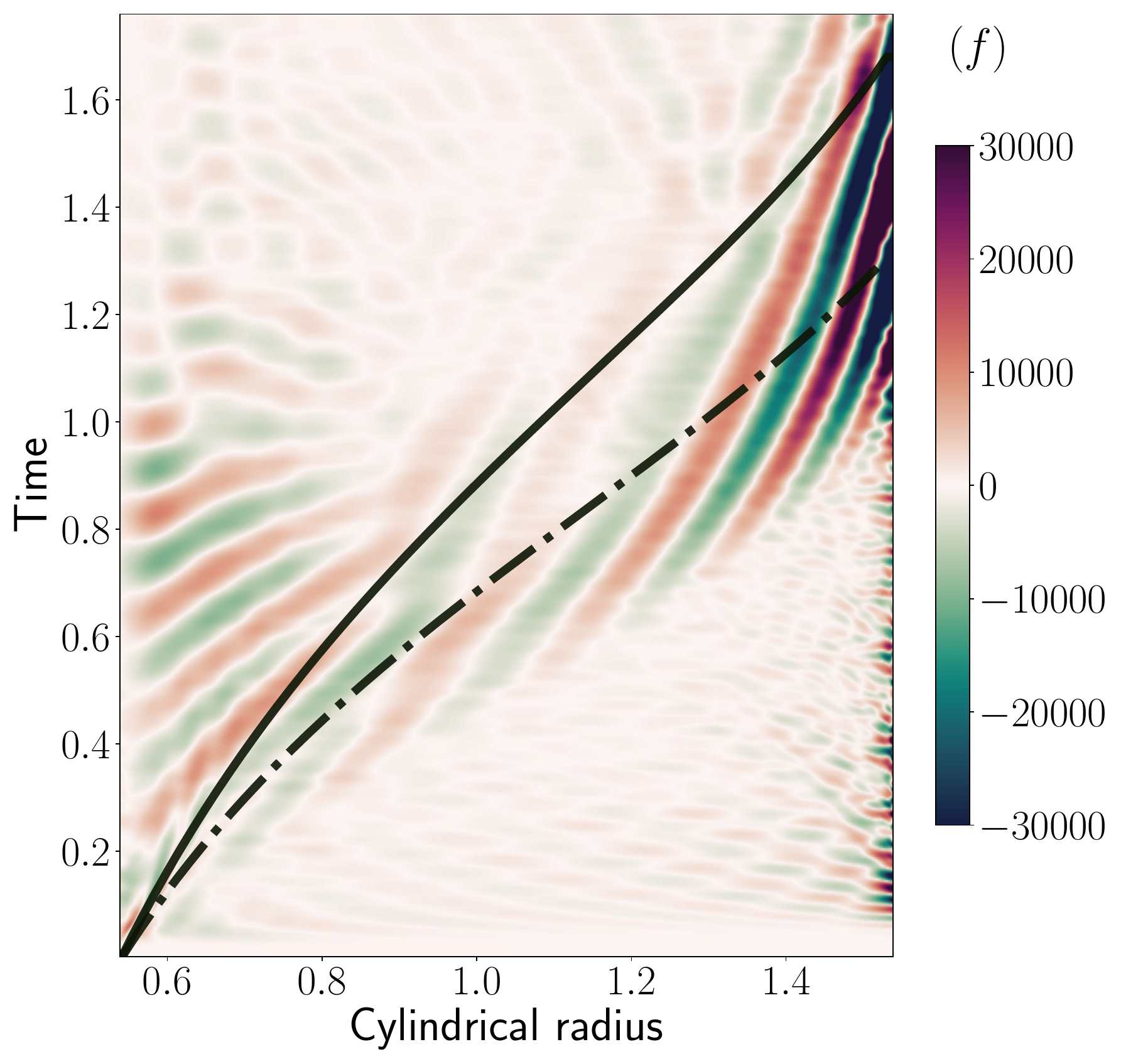}
}
	\caption{
	Temporal evolution of the $z$-averaged and curled inertia $\left< \nabla \times (\partial_t {\bm u}) \right>$ (panels a, d), Lorentz force $\left< \nabla \times (\nabla \times {\bf B} \times {\bf B}) \right> / (Pm\,\lambda)$ (panels b, e) and Coriolis force $\left< \nabla \times (-2 {\bm e}_z  \times {\bm u}) \right> / \lambda$ (panels c, f) for a case at $Ek= 10^{-7}$, $Pm=0.144$ and $S = 1596$ (panels a--c) and a case at $Ek= 3 \times 10^{-10}$, $Pm=7.9 \times 10^{-3}$ and $S = 6825$ (panels d--f).
    The force balance is taken at a particular 'fast' longitude delineated in light blue in Fig.~\ref{fig:B_background}.
    Theoretical ray-tracing trajectories at the Alfv\'en velocity corresponding to this longitude $V_{\cal A}(s, \phi)$, and at the axisymmetrically-averaged Alfv\'en velocity $\overline{V}_{\cal A}(s)$ are respectively shown in dotted black and plain black lines -- they have been obtained by solving for the eikonal equation for the varying Alfv\'en velocities.
    }
	\label{fig:Col_Forces_Case1-2}
\end{figure*}

In this section, we analyse the vorticity balance and examine the $z$-average and curl of the forces of the problem.
For the sake of simplicity however, we refer to the quantities from the vorticity balance as the forces they correspond to in the momentum equation (eq.~\ref{eq:momentum_no-T_linearised}).

Looking at the evolution of the signal along the time axis in Figure~\ref{fig:Col_Forces_Case1-2}, we can see that it is first dominated by a balance between inertia and the Lorentz force which have the highest amplitudes (bottom left corners of each plot).
There, the propagation of the signal is compatible with a wave packet travelling at the local Alfv\'en velocity at this particular longitude $V_{\cal A}(s, \phi)$ (and not the longitudinally-averaged Alfv\'en velocity).
After about an Alfv\'en time (about $t \sim 0.7 \tau_{\cal A}$) and in the middle of the bulk (at $s \sim 1.1$) the inertial term fades out while the Coriolis force takes over.
The signal is then dominated by a balance between the Lorentz and the Coriolis forces up to the outer core boundary (top right corner of each plot), and propagation is slower under this force balance.

Therefore, the waves see their force balance and nature changing as they propagate outwards: QG-Alfv\'en waves -- characterised by a balance between the Lorentz force and inertia -- are emitted following the initial perturbation and transform into slower QG-Magneto-Coriolis waves -- characterised by a balance between the Lorentz force and the Coriolis force -- while approaching the core mantle boundary under the effect of the container curvature -- increasing the predominance of Coriolis.
The decreasing velocity of the QG-MC wave front is consistent with the fact that these waves are dispersive contrary to the QG-Alfv\'en waves (see Table~\ref{tab:wave_types}).
Moreover, we can observe in both the top and bottom panels that the period (waves length-scale along the time axis) remains mostly unchanged while the QG-Alfv\'en wave front becomes QG-MC.

Additionally in Fig.~\ref{fig:Col_Forces_Case1-2}, Rossby waves -- characterised by a balance between inertia and the Coriolis force -- are visible in inertia and the Coriolis force shortly after the impulse (rather faint signal at the bottom of inertia and Coriolis plots).
And we can also observe that inward waves are re-emitted at the CMB after the Rossby waves have reached it.
Some of the re-emitted waves are characterised by a balance between the Lorentz force and inertia, indicating inward QG-Alfv\'en waves, while other waves are characterised by a balance between the Coriolis force and inertia, indicating inward Rossby waves. 
It appears that Rossby waves, triggered by the impulse at the start of the experiment, provoke the re-emission of inward Rossby and QG-Alfv\'en waves -- probably by non-linear wave-wave interactions -- when they reach the CMB after a short delay (about $t \sim 0.07 \tau_{\cal A}$), well before the arrival of the main wave front.
This proposed scenario is also compatible with what was observed in the equatorial cross-sections shown in Fig.~\ref{fig:Maps_t_evolution}~(a-b).
Note that we have observed no major differences for the waves and their dynamics when varying the strength of the hyperdiffusivity for the case at $Ek = 10^{-7}$ (not shown), indicating that the large length-scales are converged and that the small length-scales probably do not play a significant role during the morphing process.

The aforementioned sequence seems independent of the parameters. 
For both cases, we first observe the emission of QG-Alfv\'en waves -- when the signal is dominated by inertia and the Lorentz force -- that progressively become QG-Magneto-Coriolis waves during the crossing of the shell -- when the signal is dominated by the Coriolis and the Lorentz forces.
And in both cases, we see that the QG-Aflv\'en front starts following the local Alfv\'en speed while the subsequent QG-Magneto-Coriolis front is slower (related to the dispersive nature of the latter signal) with similar time arrivals.
The only differences seem to be that the faint Rossby waves (at the start) followed by the re-emitted inward Rossby and QG-Alfv\'en waves at the outer boundary appear slightly dimer in the lower Ekman case (bottom panel).
The tests conducted with a longer time window also show that the waves of the system reflect at the boundaries and slowly dissipate, bouncing back and forth for at least several Alfv\'en times (not shown).

Checking for both cases the rms force balance's residuals -- {\it i.e.} $z$-averaged Lorentz minus Coriolis minus $d \omega_z / dt$, which should be an estimate of the viscous force although some departure from the main equilibrium could remain in this quantity -- (not shown), we find that the strongest residuals are located along the ICB and the CMB and that they are mainly present at the start of the simulation (following the super-rotation of the inner core).
On the contrary, there are almost no residuals in the bulk and after more than an Alfv\'en time near the ICB.
Between the two cases, these residuals are greatly reduced, especially near the CMB.
This confirms that residuals in our problem are viscous -- the last remaining force of the system in the curled balance.
This additionally confirms that the balance between inertia, the Lorentz force and the Coriolis force (observed in Fig.~\ref{fig:Col_Forces_Case1-2}) is largely respected in the bulk while the viscous force only plays a marginal role in the observed wave dynamics -- and only weakening with decreasing $Pm$.
These results are characteristics of all our computations.

\subsection{Energy ratios}
\label{sec:col_spectra}

\begin{figure*}
\centering{
	\includegraphics[width=0.49\linewidth]{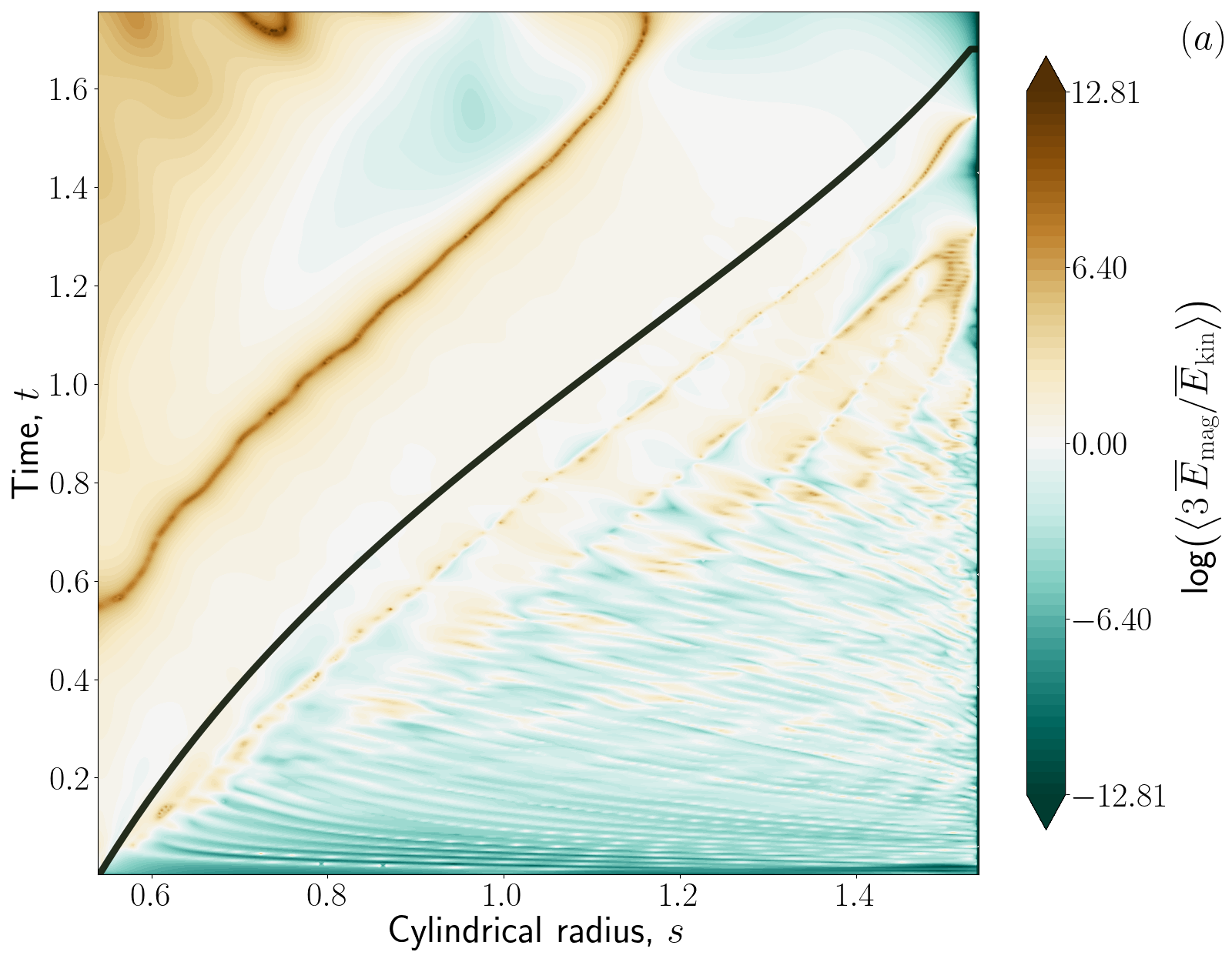}
    \includegraphics[width=0.49\linewidth]{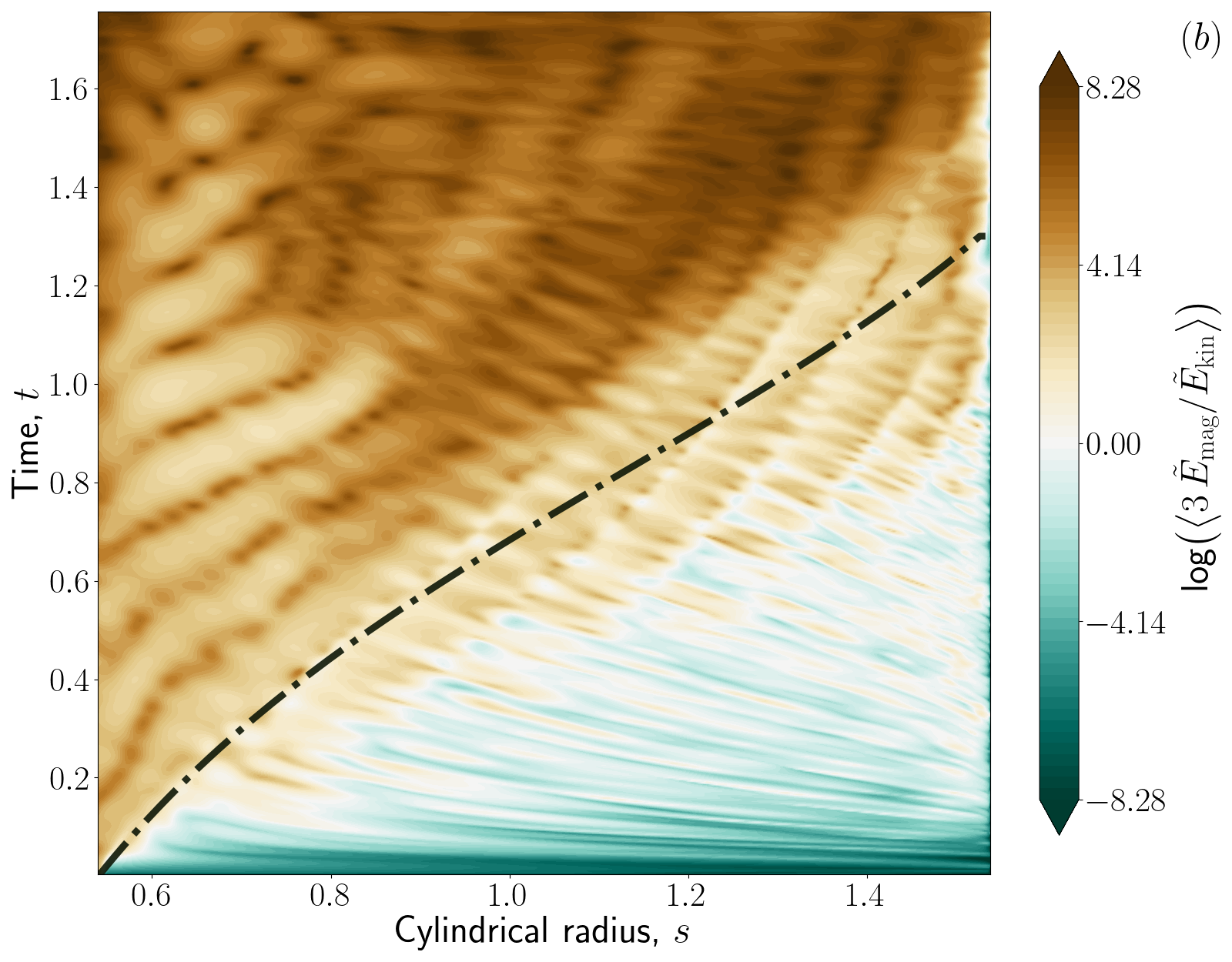}
}
	\caption{
	Temporal evolution of the $z$-and-$\phi$-averaged (zonal) magnetic to kinetic energy ratio $\mathrm{log}\left( \left< 3\,\overline{E}_\mathrm{mag} / \overline{E}_\mathrm{kin} \right> \right)$ (a) and of the $z$-averaged non-zonal magnetic to kinetic energy ratio $\mathrm{log}\left( \left< 3\,\tilde{E}_\mathrm{mag} / \tilde{E}_\mathrm{kin} \right> \right)$ (b) for a case at $Ek= 10^{-7}$, $Pm=0.144$ and $S = 1596$.
    Theoretical ray-tracing trajectories at the axisymmetrically-averaged Alfv\'en velocity $\overline{V}_{\cal A}(s)$, and the Alfv\'en velocity corresponding to this longitude $V_{\cal A}(s, \phi)$ are respectively shown with the plain black (a) and dotted black lines (b) -- they have been obtained by solving for the eikonal equation for the varying Alfv\'en velocities.
    The colormaps for the energy levels are given in log-scales.
	}
	\label{fig:Emag-Ekin-ratio}
\end{figure*}

Magnetic to kinetic energy ratios are displayed for the case at $Ek= 10^{-7}$, $Pm=0.144$ and $S = 1596$ in Figure~\ref{fig:Emag-Ekin-ratio}, and we clearly retrieve in Fig.~\ref{fig:Emag-Ekin-ratio}~(a) the torsional wave which presents an energetic equipartition ($E_\mathrm{mag} \simeq E_\mathrm{kin}$).
Note that contrary to ${\bm u}$, ${\bf b}$ is not quasi-perfectly columnar and a residual geometric factor $3$ had to be used to properly scale the $z$-averaged energy ratios.
In Fig.~\ref{fig:Emag-Ekin-ratio}~(b), we can track the evolution of the wave front (as can be seen in Fig.~\ref{fig:Col_Forces_Case1-2}~b) with QG-Alfv\'en waves associated with an energetic equipartition $E_\mathrm{mag} \simeq E_\mathrm{kin}$ at the start of the computation (bottom-left), exhibiting an increasing dominance of the magnetic energy as the wave front approaches $s = s_o$, and with QG-MC waves associated with $E_\mathrm{mag} \gg E_\mathrm{kin}$ at the end of the simulation (top-right).
As expected, the Rossby waves domain (bottom-right) is dominated by the kinetic energy and we can also see the inward QG-Alfv\'en waves where the magnetic to kinetic energy ratio is closer to $1$.
This is consistent with what could be concluded from the previous Fig.~\ref{fig:Col_Forces_Case1-2} and from the taxonomy of the waves expected in our system (see Table~\ref{tab:wave_types}).
  
\subsection{Dispersion relation}
\label{sec:disp_rel}

To better quantify our results, we will now investigate in more details the dispersion relation when the QG-Alfv\'en wave front transforms into the QG-Magneto-Coriolis wave front.

\subsubsection{Derivation}
\label{sec:disp_deriv}

Under the plane wave ansatz, assuming that the radial length-scales are much shorter than the horizontal length-scales, that the azimuthal magnetic field is not significantly larger than the radial one, and that the radial wavelength of the perturbation is much shorter than the length-scales over which the background fields evolve, while neglecting the magnetic dissipation, \cite{gillet2022satellite} derived a dispersion relation for the mixed QG-Alfv\'en-Magneto-Coriolis wave:
\begin{align}
\label{eq:disp_rel_QGA}
\omega = \dfrac{m \Omega}{k^2 h(s)^2} \pm \displaystyle\sqrt{\left( \dfrac{m \Omega}{k^2 h(s)^2} \right)^2 + V_{\cal A}(s, \phi)^2 k^2} \,,
\end{align}
where $m$ and $k$ are respectively the azimuthal and the cylindrically radial wave numbers.
Note that the theoretical ray-tracing trajectories (in {\it e.g.}, Fig.~\ref{fig:Col_Forces_Case1-2}) that are solutions for the eikonal equation for varying Alfv\'en velocities can be related to this dispersion relation.
They indicate that a wave packet propagation is consistent with the prediction for a QG-Alfv\'en wave packet.

However, because of the short and non-periodic impulse we impose at the start, our setup triggers a wave packet rather than plane waves and does not allow us to precisely control and track specific wavenumbers.
A wave packet and group velocity approach are therefore probably better suited given the nature of our experiment.

\subsubsection{Group velocity}
\label{sec:group_vel}
The group velocity can be derived from eq.~(\ref{eq:disp_rel_QGA}), and the cylindrical radial component of the group velocity then reads
\begin{align}
\label{eq:group_vel_QGA}
\dfrac{\partial \omega}{\partial k} \simeq \pm V_{\cal A}(s, \phi) - \dfrac{2 m \Omega}{k^3 h(s)^2}\,.
\end{align}
In this expression, we can see that for small radial wavelengths away from the outer boundary (large $k$, large $h$), the propagation speed should be close to the local Alfv\'en speed $V_{\cal A}$, promoting QG-Alfv\'en waves.
When approaching the outer boundary, under the increasing influence of the Coriolis force due to the topographic $\beta$-effect, the second term in (\ref{eq:group_vel_QGA}) becomes larger (smaller and smaller $h$) and the group speed decelerates.
The wave would then slow down and transition from a QG-Alfv\'en to a QG-Magneto-Coriolis wave while reaching the CMB.
The dispersive nature of the QG-MC wave under the influence of the second term of eq.~(\ref{eq:group_vel_QGA}) is revealed as small radial length-scale (large $k$) and large azimuthal length-scale (small $m$) see their speed impeded the least.

\begin{figure*}
\centering{
	\includegraphics[width=.99\linewidth]{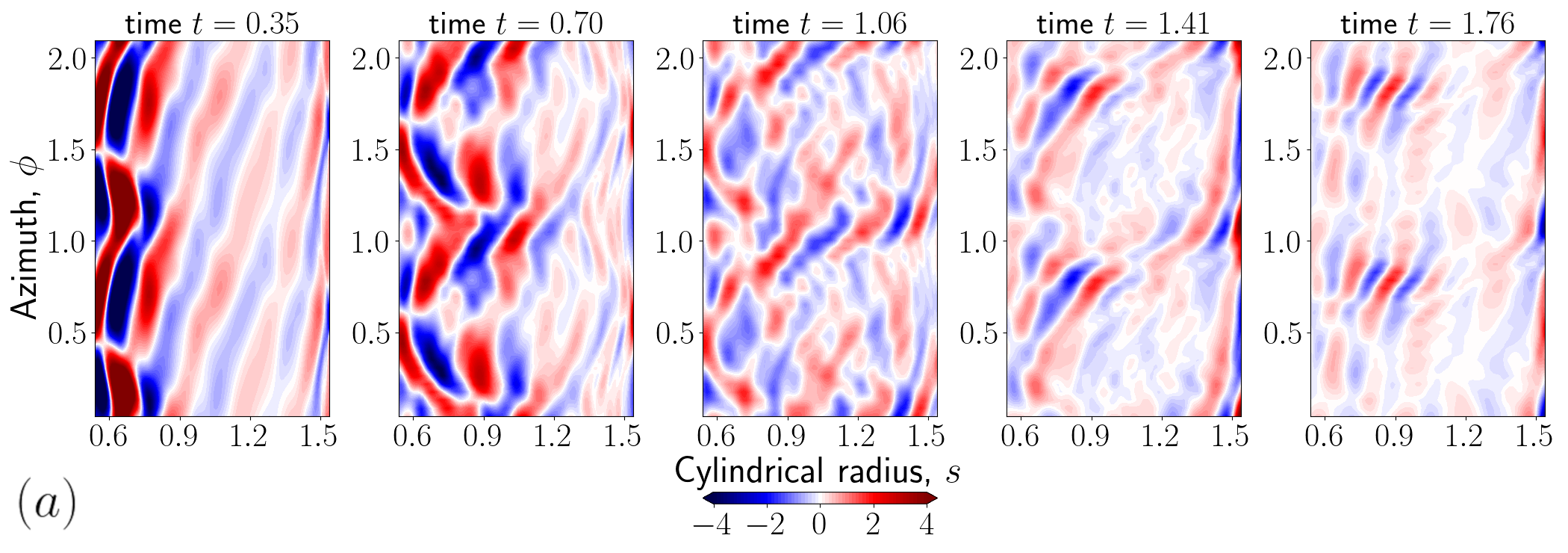}}
\centering{
	\includegraphics[width=.99\linewidth]{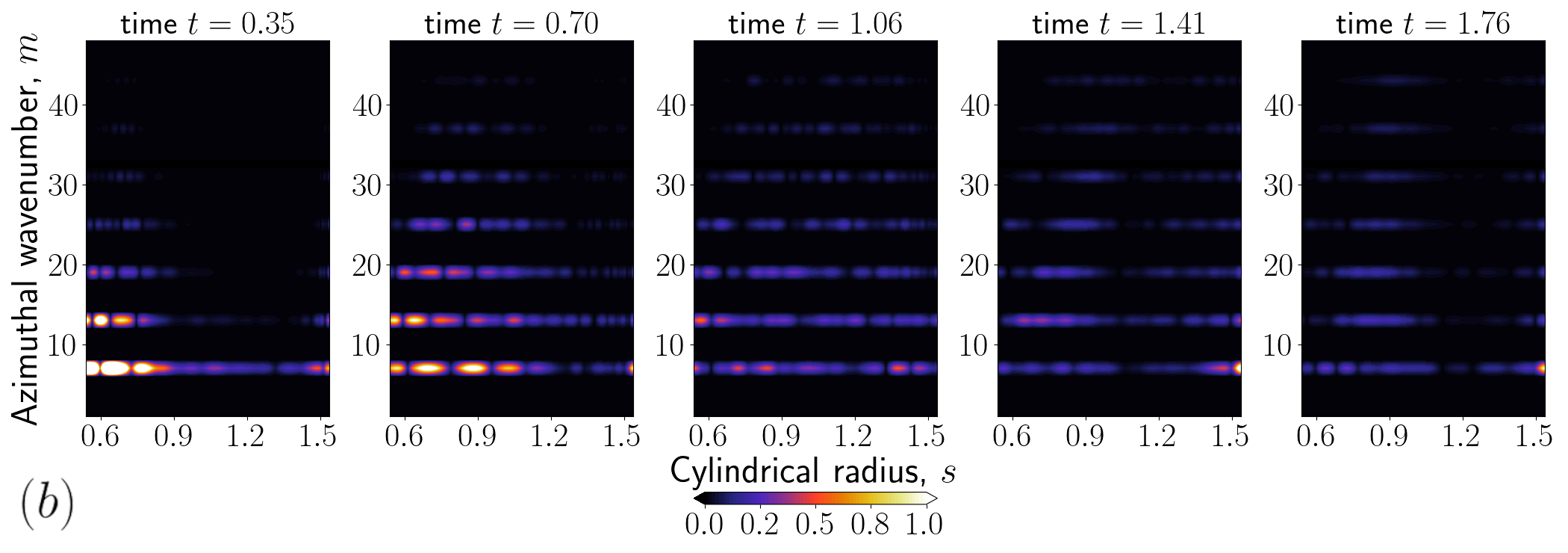}}
	\caption{
	Series of snapshots of the columnar non-zonal azimuthal velocity $\left< \tilde{u}_\phi \right>$ in the equatorial plane (a), and of the same field where the $\phi$-axis has been transformed in the spectral space using an FFT (b) for a case at $Ek= 10^{-7}$, $Pm=0.144$ and $S = 1596$.
    Note that these plots have the same $s$-axis and are plotted at the sames times.
    Times are given in terms of the Alfv\'en timescale $\tau_{\cal A}$.
	}
	\label{fig:Transformed_t_evolution}
\end{figure*}

Looking at the non-zonal velocity field for a case at $Ek= 10^{-7}$, $Pm=0.144$ and $S = 1596$ in Figure~\ref{fig:Transformed_t_evolution}, we can first see that the columnar non-zonal velocity (Fig.~\ref{fig:Transformed_t_evolution}~a) is very similar to the velocity observed in the equatorial plane (see Fig.~\ref{fig:Maps_t_evolution}~b), which is compatible with the low Lehnert numbers of our simulations ($\left< {\bm u} \right> \sim {\bm u}\vert_\mathrm{equator}$).
Then we retrieve the sequence already described in \S~\ref{sec:Time_evo} and \S~\ref{sec:col_bal} where a wave packet, localised in $s$ and $\phi$ and identified as QG-Alfv\'en waves, slowly travels outward while a faster signal, azimuthally elongated and identified as Rossby waves, rapidly reach the CMB.
In the bulk, the system becomes more complex and the signals are disturbed during the crossing.
When the slower wave packet approaches the outer boundary, it transforms into QG-MC waves, notably becoming azimuthally elongated.
During this transition from QG-Alfv\'en to QG-MC waves (that we can observe from times $t = 0.7\,\tau_{\cal A}$ to $t = 1.41\,\tau_{\cal A}$ in Fig.~\ref{fig:Transformed_t_evolution}~a), the waves become thinner in the cylindrical radial direction ($k$ increases) but wider in the azimuthal direction ($m$ decreases), suggesting that the period of the wave front remains mostly unchanged (following eq.~\ref{eq:group_vel_QGA}).

This sequence is consistent with what can be observed in the bottom panel (Fig.~\ref{fig:Transformed_t_evolution}~b).
At the start of the simulation, most of the energy is contained in the azimuthal wavenumber $m=6$ at low radii but some energy can already be observed at the CMB (bottom right corner of the plots), consistent with the observations of the top panel at the same time.
As time advances, the system is enriched with more and more azimuthal wavenumbers (see times $t = 0.7 \tau_{\cal A}$ to $t = 1.06 \tau_{\cal A}$) and the energy becomes more distributed across various $m$ while the energy maxima slowly progress outward.
Note that we retrieve a dispersion of the signal as the higher $m$ (smaller azimuthal length-scales) seem slower than the smaller $m$ on these snapshots.
Toward the end of the computation, as the energy maximum reaches the outer boundary, the complexity is progressively lost and larger $m$ fade out.
Only a remnant signal can be observed around $s \sim 0.8 s_o$ (in both panels at the same time), which can be related to inward propagating QG-Alfv\'en waves (see \S~\ref{sec:col_bal}).
Fig.~\ref{fig:Transformed_t_evolution}~(b) shows that all the retrieved azimuthal wavenumbers seem to be multiples of $m = 6$ (consistent with observations that could already be made in Fig.~\ref{fig:Maps_t_evolution}~b or  Fig.~\ref{fig:Transformed_t_evolution}~a).
This is because of the $m = 3$ background magnetic field symmetry as the waves depend on the geometry of $B_0^2$, implying that there are two fast directions per sub-domain (see Fig.~\ref{fig:B_background}), and thus promoting the $m = 6$ family.

Overall, we retrieve here a system that can be understood as a packet of QG-Alfv\'en waves which progressively becomes a packet of QG-Magneto-Coriolis waves under the increasing influence of the Coriolis force while approaching the outer boundary.
Rossby waves and inward QG-Alfv\'en waves (that can also be spotted in Fig.~\ref{fig:Transformed_t_evolution}~a) complete the system.

\section{Summary and discussion}
\label{sec:Conclusion}

This work demonstrates that a complex system of QG waves develops over a non-axisymmetric magnetic background following a rotational impulse of the inner core.
After the initial impulse, a packet of QG-Alfv\'en waves is one of the first type of waves to be emitted.
This wave packet evolves into QG-Magneto-Coriolis waves when approaching the outer core boundary under the growing influence of the Coriolis force.
Besides these two types of waves, a torsional Alfv\'en wave and 'slow' Rossby waves -- both triggered by the initial impulse -- are also observed.
A complex mixture of inward Rossby and QG-Aflv\'en waves -- re-emitted at the CMB after the arrival of the initial Rossby waves -- complement the rich dynamics of this system.
Residuals are mainly viscous in nature and tend to disappear when lowering the Ekman and magnetic Prandtl numbers.
The analysis of the vorticity force balance, of the energy ratios and of the group velocity have consolidated our interpretation of the system's dynamics.

The torsional Alfv\'en wave retrieved here is produced by complex interactions in the shear layer at the inner core boundary because of the nature of our impulse \cite{jault2008axial}.
It is not clear whether or not there are interactions between the torsional Alfv\'en wave and the QG-MC waves \cite{labbe2015magnetostrophic} and the torsional wave remains mostly independent of the other signals in our configuration.
Nonetheless, we have revisited a previously proposed scaling law for the thickness of the torsional Alfv\'en wave \cite{jault2008axial}, and found here that its thickness scales as $Ek_M^{1/4}$.

Concerning the non-axisymmetric signals, we find that disruptions in the underlying $1^\mathrm{st}$ order balance first involve inertia and causes the initial propagation of QG-Alfv\'en waves -- under an equilibrium between inertia and the Lorentz force -- as well as Rossby waves -- under an equilibrium between inertia and the Coriolis force. 
The QG-Alfv\'en waves then evolve into QG-Magneto-Coriolis waves because of the growing predominance of the Coriolis force when approaching the CMB.
The QG-MC wave front mostly conserves its period but shrinks in the radial direction and becomes azimuthally elongated as QG-MC waves are dispersive contrary to QG-Alfv\'en waves.
Similarly to \cite{gerick2021fast}, we found that QG-MC waves do not appear to be strongly influenced by the diffusion, or at least if the timescale of the diffusion is sufficiently larger than the other timescales (large value of $S$).
This sequence is similar to what have been found in a full 3D setup with time-dependent background magnetic field \cite{aubert2022taxonomy}.
The QG-MC wave front is visible near the core surface around $1\,\tau_{\cal A}$, at the earliest, after the initial perturbation, that is about $\sim 2\,y$ for the Earth.
More or less independently of the parameters, we find that when reaching the CMB, the QG-MC waves have an estimated outward propagation $V_s \sim 280\,km/y$, and display a clear westward drift with an estimated phase speed $V_\phi \sim 1100\,km/y$.
These values are compatible with observation-based estimates of QG-MC velocities, with $V_s \sim 200\,km/y$ and $V_\phi \sim 1500\,km/y$ reported by \cite{gillet2022satellite}.
This confirms the QG-MC nature of the interannual magnetic signals observed near the equator (in agreement with \cite{gerick2021fast,gillet2022satellite,aubert2022taxonomy}).

The limitations of our study are two folds.
First, the role of small length-scales in the reported dynamics can be questioned.
Whilst the small length-scales are likely not instrumental during the morphing process of the QG-Alfv\'en into QG-MC waves -- as the large length-scales and the waves dynamics remain unchanged even when substantially relaxing the hyperdiffusivity -- they probably affect the reflection of the waves at the boundaries.
This is, however, considered beyond the scope of this work as we focus here on the transient dynamics of the system.
We also reach sufficiently low Ekman numbers to think that the effect of the viscous boundary layers and the velocity boundary conditions would not significantly impact the reported results.
Then, the influence of our particular configuration can be interrogated.
Our background magnetic field has no toroidal component and the imposed $m = 3$ symmetry only triggers waves with limited azimuthal wavenumbers. 
In a recent study exploring the impact of the background magnetic field on QG-MC modes -- including poloidal and toroidal components --, it has been found that 'geomagnetically relevant' QG-MC modes do not seem strongly affected by the toroidal component of the background magnetic field and that they are supported by a variety of magnetic base states as long as those are non-axisymmetric \cite{gerick2024interannual}.
Hence, we do not expect the geometry of the background magnetic field to drastically change the morphing process of the QG-MC waves, nor to impede their occurrence near the core surface, but it can certainly impact the geophysical relevance of the retrieved QG-MC waves.

Finally, several avenues can be thought of in order to improve on the current analysis.
Here, a single rotational pulse of the inner core triggers the initial wave packets and the influence of the forcing on the waves that are emitted remains to be analysed.
For example, periodic or stochastic impulses would generate a succession of waves that might make the study of a dispersion relation easier.
Different kinds of impulses could also be tested -- {\it e.g.}, a velocity burst in the bulk of the fluid -- to examine the dependence of the wave geometries on the nature of the forcing.
More diverse background magnetic fields, with higher azimuthal symmetries, or even departing from simple azimuthal geometries -- such as outputs from data assimilation \cite{aubert2023state} -- could also be tested to study the QG-wave dynamics in a more realistic setup.
In this latter case, the effect of the magnetic diffusion might be studied in more details as it has been suggested to have an impact on transient QG-Alfv\'en waves in complex background magnetic fields \cite{aubert2021interplay}. 

The long-term goal remains to characterise QG-Alfv\'en and QG-MC waves in more complex setups and to apply this knowledge to Earth's geomagnetic observations. 
This reduced model could also be used in the future as a basis for data-assimilation targeting wave-like pattern in the geomagnetic signal.

\section*{Acknowledgements}
We thank two anonymous reviewers and Bruce Buffett for their comments and suggestions that greatly helped improve our manuscript.
This project has been funded by ESA in the framework of EO Science for Society, through contract 4000127193/19/NL/IA (SWARM + 4D Deep Earth: Core). Numerical computations were performed at S-CAPAD, IPGP.

\section*{Data availability}

Additional data are provided and available at \url{https://doi.org/10.5281/zenodo.10609678}.
The python package {\tt parobpy} and the scripts used to produce the results shown in this manuscript are available at \url{https://github.com/OBarrois/parobpy} (DOI: \url{10.5281/zenodo.12820318}).

\bibliography{artbib}

\begin{thebibliography}{99}

\bibitem{courtillot1978acceleration}
Courtillot V. 1978  Sur une acc{\'e}l{\'e}ration r{\'e}cente de la variation
  s{\'e}culaire du champ magn{\'e}tique terrestre. {\em CR Hebd. SEances Acad.
  Sci. Paris} pp. 1095--1098.

\bibitem{chulliat2010observation}
Chulliat A, Olsen N. 2010  Observation of magnetic diffusion in the {E}arth's
  outer core from {M}agsat, {{\O}}rsted, and {CHAMP} data. {\em J. Geophys.
  Res.: Solid Earth} \textbf{115}, 1--13.

\bibitem{pinheiro2011measurements}
Pinheiro K, Jackson A, Finlay C. 2011  Measurements and uncertainties of the
  occurrence time of the 1969, 1978, 1991, and 1999 geomagnetic jerks. {\em
  Geochemistry, Geophysics, Geosystems} \textbf{12}.

\bibitem{brown2013jerks}
Brown W, Mound J, Livermore P. 2013  Jerks abound: An analysis of geomagnetic
  observatory data from 1957 to 2008. {\em Physics of the Earth and Planetary
  Interiors} \textbf{223}, 62--76.

\bibitem{finlay2020chaos}
Finlay CC, Kloss C, Olsen N, Hammer MD, T{\o}ffner-Clausen L, Grayver A,
  Kuvshinov A. 2020  The CHAOS-7 geomagnetic field model and observed changes
  in the South Atlantic Anomaly. {\em Earth, Planets and Space} \textbf{72},
  1--31.

\bibitem{lesur2010modelling}
Lesur V, Wardinski I, Asari S, Minchev B, Mandea M. 2010  Modelling the
  {E}arth’s core magnetic field under flow constraints. {\em Earth, planets
  and space} \textbf{62}, 503--516.

\bibitem{finlay2016recent}
Finlay CC, Olsen N, Kotsiaros S, Gillet N, T{\o}ffner-Clausen L. 2016  Recent
  geomagnetic secular variation from {S}warm. {\em Earth, Planets and Space}
  \textbf{68}, 1--18.

\bibitem{chulliat2014geomagnetic}
Chulliat A, Maus S. 2014  Geomagnetic secular acceleration, jerks, and a
  localized standing wave at the core surface from 2000 to 2010. {\em J.
  Geophys. Res.: Solid Earth} \textbf{119}, 1531--1543.

\bibitem{chulliat2015fast}
Chulliat A, Alken P, Maus S. 2015  Fast equatorial waves propagating at the top
  of the {E}arth's core. {\em Geophys. Res. Lett.} \textbf{42}, 3321--3329.

\bibitem{aubert2019geomagnetic}
Aubert J, Finlay CC. 2019  Geomagnetic jerks and rapid hydromagnetic waves
  focusing at Earth’s core surface. {\em Nature Geoscience} \textbf{12},
  393--398.

\bibitem{aubert2022taxonomy}
Aubert J, Livermore PW, Finlay CC, Fournier A, Gillet N. 2022  A taxonomy of
  simulated geomagnetic jerks. {\em Geophysical Journal International}
  \textbf{231}, 650--672.

\bibitem{aubert2021interplay}
Aubert J, Gillet N. 2021  The interplay of fast waves and slow convection in
  geodynamo simulations nearing Earth’s core conditions. {\em Geophysical
  Journal International} \textbf{225}, 1854--1873.

\bibitem{gerick2021fast}
Gerick F, Jault D, Noir J. 2021  Fast quasi-geostrophic Magneto-Coriolis modes
  in the Earth's core. {\em Geophysical Research Letters} \textbf{48},
  e2020GL090803.

\bibitem{davidson2013scaling}
Davidson P. 2013  Scaling laws for planetary dynamos. {\em Geophysical Journal
  International} \textbf{195}, 67--74.

\bibitem{roberts1965analysis}
Roberts P, Scott S. 1965  On analysis of the secular variation. {\em Journal of
  geomagnetism and geoelectricity} \textbf{17}, 137--151.

\bibitem{dormy2016strong}
Dormy E. 2016  Strong-field spherical dynamos. {\em Journal of Fluid Mechanics}
  \textbf{789}, 500--513.

\bibitem{schaeffer2017turbulent}
Schaeffer N, Jault D, Nataf HC, Fournier A. 2017  Turbulent geodynamo
  simulations: a leap towards {E}arth's core. {\em Geophys. J. Int.}
  \textbf{211}, 1--29.

\bibitem{aubert2019approaching}
Aubert J. 2019  Approaching Earth’s core conditions in high-resolution
  geodynamo simulations. {\em Geophysical Journal International} \textbf{219},
  S137--S151.

\bibitem{schwaiger2019force}
Schwaiger T, Gastine T, Aubert J. 2019  Force balance in numerical geodynamo
  simulations: a systematic study. {\em Geophysical Journal International}
  \textbf{219}, S101--S114.

\bibitem{zhang2001inertial}
Zhang K, Earnshaw P, Liao X, Busse FH. 2001  On inertial waves in a rotating
  fluid sphere. {\em Journal of Fluid Mechanics} \textbf{437}, 103--119.

\bibitem{canet2014hydromagnetic}
Canet E, Finlay CC, Fournier A. 2014  Hydromagnetic quasi-geostrophic modes in
  rapidly rotating planetary cores. {\em Physics of the Earth and Planetary
  Interiors} \textbf{229}, 1--15.

\bibitem{buffett2019equatorially}
Buffett B, Matsui H. 2019  Equatorially trapped waves in Earth’s core. {\em
  Geophysical Journal International} \textbf{218}, 1210--1225.

\bibitem{aubert2023state}
Aubert J. 2023  State and evolution of the geodynamo from numerical models
  reaching the physical conditions of Earth’s core. {\em Geophysical Journal
  International} \textbf{235}, 468--487.

\bibitem{lesur2022rapid}
Lesur V, Gillet N, Hammer M, Mandea M. 2022  Rapid variations of Earth’s core
  magnetic field. {\em Surveys in Geophysics} \textbf{43}, 41--69.

\bibitem{gillet2010fast}
Gillet N, Jault D, Canet E, Fournier A. 2010  Fast torsional waves and strong
  magnetic field within the {E}arth’s core. {\em Nature} \textbf{465},
  74--77.

\bibitem{labbe2015magnetostrophic}
Labb{\'e} F, Jault D, Gillet N. 2015  On magnetostrophic inertia-less waves in
  quasi-geostrophic models of planetary cores. {\em Geophysical \&
  Astrophysical Fluid Dynamics} \textbf{109}, 587--610.

\bibitem{gillet2022satellite}
Gillet N, Gerick F, Jault D, Schwaiger T, Aubert J, Istas M. 2022  Satellite
  magnetic data reveal interannual waves in Earth’s core. {\em Proceedings of
  the National Academy of Sciences} \textbf{119}, e2115258119.

\bibitem{jault2008axial}
Jault D. 2008  Axial invariance of rapidly varying diffusionless motions in the
  Earth’s core interior. {\em Physics of the Earth and Planetary Interiors}
  \textbf{166}, 67--76.

\bibitem{gillet2011rationale}
Gillet N, Schaeffer N, Jault D. 2011  Rationale and geophysical evidence for
  quasi-geostrophic rapid dynamics within the {E}arth's outer core. {\em Phys.
  Earth Planet. Int.} \textbf{187}, 380--390.

\bibitem{de1998viscosity}
De~Wijs GA, Kresse G, Vo{\v{c}}adlo L, Dobson D, Alfe D, Gillan MJ, Price GD.
  1998  The viscosity of liquid iron at the physical conditions of the Earth's
  core. {\em Nature} \textbf{392}, 805--807.

\bibitem{pozzo2014thermal}
Pozzo M, Davies C, Gubbins D, Alf{\`e} D. 2014  Thermal and electrical
  conductivity of solid iron and iron--silicon mixtures at Earth's core
  conditions. {\em Earth and Planetary Science Letters} \textbf{393}, 159--164.

\bibitem{gillet2015planetary}
Gillet N, Jault D, Finlay C. 2015  Planetary gyre, time-dependent eddies,
  torsional waves, and equatorial jets at the {E}arth's core surface. {\em J.
  Geophys. Res.: Solid Earth} \textbf{120}, 3991--4013.

\bibitem{baerenzung2018modeling}
B{\"a}renzung J, Holschneider M, Wicht J, Sanchez S, Lesur V. 2018  Modeling
  and Predicting the Short-Term Evolution of the Geomagnetic Field. {\em
  Journal of Geophysical Research: Solid Earth} \textbf{123}, 4539--4560.

\bibitem{aubert2013bottom}
Aubert J, Finlay CC, Fournier A. 2013  Bottom-up control of geomagnetic secular
  variation by the {E}arth's inner core. {\em Nature} \textbf{502}, 219--223.

\bibitem{aubert2017spherical}
Aubert J, Gastine T, Fournier A. 2017  Spherical convective dynamos in the
  rapidly rotating asymptotic regime. {\em Journal of Fluid Mechanics}
  \textbf{813}, 558--593.

\bibitem{schaeffer2013efficient}
Schaeffer N. 2013  Efficient spherical harmonic transforms aimed at
  pseudospectral numerical simulations. {\em Geochemistry, Geophysics,
  Geosystems} \textbf{14}, 751--758.

\bibitem{nataf2015turbulence}
Nataf HC, Schaeffer N. 2015  Turbulence in the core. .

\bibitem{gillet2022dynamical}
Gillet N, Gerick F, Angappan R, Jault D. 2022  A dynamical prospective on
  interannual geomagnetic field changes. {\em Surveys in Geophysics}
  \textbf{43}, 71--105.

\bibitem{braginsky1970torsional}
Braginsky S. 1970  Torsional magnetohydrodynamic vibrations in the Earth's core
  and variations in day length. {\em Geomagn. Aeron.} \textbf{10}, 3--12.

\bibitem{zhang1993equatorially}
Zhang K. 1993  On equatorially trapped boundary inertial waves. {\em Journal of
  Fluid Mechanics} \textbf{248}, 203--217.

\bibitem{hide1966free}
Hide R. 1966  Free hydromagnetic oscillations of the Earth's core and the
  theory of the geomagnetic secular variation. {\em Philosophical Transactions
  of the Royal Society of London. Series A, Mathematical and Physical Sciences}
  \textbf{259}, 615--647.

\bibitem{malkus1967hydromagnetic}
Malkus WV. 1967  Hydromagnetic planetary waves. {\em Journal of Fluid
  Mechanics} \textbf{28}, 793--802.

\bibitem{gerick2024interannual}
Gerick F, Livermore P. 2024  Interannual Magneto-Coriolis modes and their
  sensitivity on the magnetic field within the Earth's core. {\em arXiv
  preprint arXiv:2403.03011}.

\end{thebibliography}

\bibliographystyle{RS.bst}

\onecolumn

\appendix
\section{Results of numerical simulations}
\label{sec:Append-A-Results}

\begin{longtable}{p{0.12\textwidth} p{0.13\textwidth} p{0.13\textwidth} p{0.04\textwidth} p{0.13\textwidth} p{0.04\textwidth} p{0.08\textwidth} p{0.04\textwidth} p{0.10\textwidth}}
\caption{Summary of the numerical simulations computed in this study using $\eta = r_i/r_o = 0.35$ as suited for the Earth's core. $Ek$ is the Ekman number, $Pm$ is the magnetic Prandtl number, $Ek_M = Ek / Pm$ is the magnetic Ekman number, $S$ is the Lundquist number, $\lambda$ is the Lehnert number, $\Lambda$ is the Elsasser number of the background magnetic field, $\ell_H$ is the cut-off harmonic degree above which the hyperdiffusion has no effect, $q_H$ is the strength of the hyperdiffusion applied to the simulation, $\delta_\mathrm{TW}$ is the width of the torsional wave, and $(N_r, \ell_{\mathrm{max}})$ are the grid-size for each of the runs.
}
\label{tab:run_list} \\
\hline
$Ek$ & $Pm$ & $Ek_M$ & $S$ & $\lambda$ & $\Lambda$ & $\ell_H / q_H$ & $\delta_\mathrm{TW}$ & $(N_r, \ell_{\mathrm{max}})$ \\
\hline
\endfirsthead

\hline
$Ek$ & $Pm$ & $Ek_M$ & $S$ & $\lambda$ & $\Lambda$ & $\ell_H / q_H$ & $\delta_\mathrm{TW}$ & $(N_r, \ell_{\mathrm{max}})$ \\
\hline
\endhead

\hline
\multicolumn{8}{c}{Continued on next page $\ldots$} \\
\hline
\endfoot

\hline
\hline
\endlastfoot

$3 \times 10^{-7}$  & $2.5 \times 10^{-1}$  & $1.2 \times 10^{-6}$ & $1214$ & $1.46 \times 10^{-3}$ & $1.77$ & $30/ 1.045$ & $0.460$ & $(370, 133)$ \\
$2 \times 10^{-7}$  & $1.44 \times 10^{-1}$  & $1.39 \times 10^{-6}$ & $798$ & $1.11 \times 10^{-3}$ & $0.88$ & $30/ 1.05$ & $0.464$ & $(450, 133)$ \\
$2 \times 10^{-7}$  & $1.44 \times 10^{-1}$  & $1.39 \times 10^{-6}$ & $1129$ & $1.57 \times 10^{-3}$ & $1.77$ & $30/ 1.05$ & $0.469$ & $(450, 133)$ \\
$2 \times 10^{-7}$  & $2.88 \times 10^{-1}$  & $6.94 \times 10^{-7}$ & $1596$ & $1.11 \times 10^{-3}$ & $1.77$ & $30/ 1.05$ & $0.422$ & $(450, 133)$ \\
$2 \times 10^{-7}$  & $1.44 \times 10^{-1}$  & $1.39 \times 10^{-6}$ & $1596$ & $2.22 \times 10^{-3}$ & $3.54$ & $30/ 1.05$ & $0.484$ & $(450, 133)$ \\
$1 \times 10^{-7}$  & $1.44 \times 10^{-2}$  & $6.94 \times 10^{-6}$ & $505$ & $3.50 \times 10^{-3}$ & $1.77$ & $30/ 1.05$ & $0.561$ & $(450, 133)$ \\
$1 \times 10^{-7}$  & $7.2 \times 10^{-2}$  & $1.39 \times 10^{-6}$ & $1009$ & $1.40 \times 10^{-3}$ & $0.88$ & $30/ 1.05$ & $0.464$ & $(450, 133)$ \\
$1 \times 10^{-7}$  & $7.2 \times 10^{-2}$  & $1.39 \times 10^{-6}$ & $1129$ & $1.57 \times 10^{-3}$ & $1.77$ & $30/ 1.05$ & $0.471$ & $(450, 133)$ \\
$1 \times 10^{-7}$  & $1.44 \times 10^{-1}$  & $6.94 \times 10^{-7}$ & $1596$ & $1.11 \times 10^{-3}$ & $1.77$ & $30/ 1.05$ & $0.424$ & $(450, 133)$ \\
$1 \times 10^{-7}$  & $1.44 \times 10^{-1}$  & $6.94 \times 10^{-7}$ & $1596$ & $1.11 \times 10^{-3}$ & $1.77$ & $30/1.025$ & $0.424$ & $(450, 133)$ \\
$1 \times 10^{-7}$  & $1.44 \times 10^{-1}$  & $6.94 \times 10^{-7}$ & $1596$ & $1.11 \times 10^{-3}$ & $1.77$ & $60/1.05$ & $0.425$ & $(450, 133)$ \\
$1 \times 10^{-7}$  & $1.44 \times 10^{-1}$  & $6.94 \times 10^{-7}$ & $1596$ & $1.11 \times 10^{-3}$ & $1.77$ & $100/1.03$ & $0.424$ & $(450, 133)$ \\
$7 \times 10^{-8}$  & $1.0 \times 10^{-1}$  & $7.0 \times 10^{-7}$ & $1590$ & $1.11 \times 10^{-3}$ & $1.77$ & $30/ 1.05$ & $0.426$ & $(450, 133)$ \\
$5 \times 10^{-8}$  & $1.44 \times 10^{-1}$  & $3.47 \times 10^{-7}$ & $1596$ & $5.54 \times 10^{-4}$ & $0.88$ & $30/ 1.05$ & $0.393$ & $(450, 133)$ \\
$5 \times 10^{-8}$  & $1.44 \times 10^{-1}$  & $3.47 \times 10^{-7}$ & $2257$ & $7.84 \times 10^{-4}$ & $1.77$ & $30/ 1.05$ & $0.398$ & $(450, 133)$ \\
$5 \times 10^{-8}$  & $1.44 \times 10^{-1}$  & $3.47 \times 10^{-7}$ & $3192$ & $1.11 \times 10^{-3}$ & $3.54$ & $30/ 1.05$ & $0.402$ & $(450, 133)$ \\
$2 \times 10^{-8}$  & $1.44 \times 10^{-1}$  & $1.39 \times 10^{-7}$ & $1596$ & $2.22 \times 10^{-4}$ & $0.35$ & $30/ 1.05$ & $0.368$ & $(450, 133)$ \\
$2 \times 10^{-8}$  & $1.44 \times 10^{-1}$  & $1.39 \times 10^{-7}$ & $7980$ & $1.11 \times 10^{-3}$ & $8.84$ & $30/ 1.05$ & $0.384$ & $(450, 133)$ \\
$1 \times 10^{-8}$  & $4.6 \times 10^{-2}$  & $2.17 \times 10^{-7}$ & $2853$ & $6.20 \times 10^{-4}$ & $1.77$ & $30/ 1.06$ & $0.369$ & $(725, 133)$ \\
$6.3 \times 10^{-9}$  & $3.65 \times 10^{-2}$  & $1.73 \times 10^{-7}$ & $3201$ & $5.53 \times 10^{-4}$ & $1.77$ & $30/ 1.05$ & $0.364$ & $(450, 133)$ \\
$6.3 \times 10^{-9}$  & $3.65 \times 10^{-2}$  & $1.73 \times 10^{-7}$ & $3201$ & $5.53 \times 10^{-4}$ & $1.77$ & $30/ 1.06$ & $0.363$ & $(725, 133)$ \\
$6.3 \times 10^{-9}$  & $3.65 \times 10^{-2}$  & $1.73 \times 10^{-7}$ & $3201$ & $5.53 \times 10^{-4}$ & $1.77$ & $30/ 1.07$ & $0.360$ & $(1025, 133)$ \\
$3 \times 10^{-10}$  & $7.9 \times 10^{-3}$  & $3.80 \times 10^{-8}$ & $6825$ & $2.59 \times 10^{-4}$ & $1.77$ & $30/ 1.09$ & $0.301$ & $(1298, 170)$
\end{longtable}

\label{lastpage}

\end{document}